\documentclass[aps,prx,twocolumn,amsmath,amssymb,10pt,aps,longbibliography,superscriptaddress,citeautoscript,bibnotes,footinbib]{revtex4-2}
\usepackage[english]{babel}
\usepackage{draftcopy}
\usepackage{graphicx}
\usepackage{verbatim}
\usepackage{color} 
\usepackage{ulem}
\usepackage{amsmath}
\usepackage{algorithm}
\usepackage{algpseudocode}
\usepackage{mathrsfs}
\usepackage{makecell}
\usepackage{algorithm,algcompatible,amsmath,mathtools}
\algnewcommand\INPUT{\item[\textbf{Input:}]}%
\algnewcommand\OUTPUT{\item[\textbf{Output:}]}%
\algnewcommand\RETURN{\item[\textbf{return}]}%

\usepackage{amsmath}
\usepackage{bm}
\usepackage{latexsym}
\usepackage{graphics}
\usepackage{dsfont}
\usepackage{amssymb}
\usepackage{tabularx}
\usepackage{bbm}
\usepackage{bbold}
\usepackage{tikz}
\usetikzlibrary{quantikz}
\usepackage[colorlinks=true]{hyperref}   
\usepackage[dvipsnames]{xcolor}

\begin{document}
\title{Adaptive sampling based quantum simulation}

\author{Sangjin Lee}
\thanks{Electronic Address: sangjin5190@gmail.com}
\affiliation{Quantum Universe Center, Korea Institute for Advanced Study, Seoul 02455, South Korea}

\author{Sangkook Choi}
\thanks{Electronic Address: sangkookchoi@kias.re.kr}
\affiliation{School of Computational Sciences, Korea Institute for Advanced Study}
\affiliation{Quantum Universe Center, Korea Institute for Advanced Study, Seoul 02455, South Korea}

\date{\today}
 
\begin{abstract}
  Early fault-tolerant quantum computers are expected to support reliable but depth-limited quantum circuits, while classical computational resources remain available. These conditions have motivated hybrid coherent algorithms which use quantum simulation as a central algorithmic primitive. This trend calls for quantum-simulation methods that operate with shallow circuits and admit systematic improvements in gate-complexity scaling. Here, we introduce qSHIFT, an adaptive sampling protocol for simulating a Hamiltonian $H=\sum_{i=1}^{L}h_iH_i$. qSHIFT achieves gate complexity $\mathcal{O}_r\left((\lambda t)^{1+1/r}/\varepsilon^{1/r}\right)$, where $r$ is an algorithmic parameter, $\lambda=\sum_i |h_i|$ and $\varepsilon$ denotes the target precision.
Relative to qDRIFT, increasing $r$ systematically improves the gate complexity for a given target precision without incurring extra quantum cost. Unlike Trotterization, the number of sampled gates is nominally independent of $L$. qSHIFT retains the elementary gate set of qDRIFT and, unlike qSWIFT, requires neither ancillary qubits nor controlled operations. The improved gate complexity scaling is obtained at the cost of a classical calculation involving $L^r$ coefficients at each adaptive sampling round.
\end{abstract}

\maketitle

\section{introduction}

Quantum simulation stands as a cornerstone application of quantum computing, offering a path to scientific breakthroughs that are computationally inaccessible to classical computers. The primary advantage of quantum computers lies in their ability to naturally represent and manipulate quantum states within an exponentially large Hilbert space\cite{qsim_general, qsim_genengal2, qsim_general3}—a task that remains the central bottleneck for classical simulations as particle numbers increase. These capabilities have profound implications for condensed matter physics~\cite{qsim_cond1, qsim_cond2,qsim_cond3,qsim_cond4,qsim_cond5,qsim_cond6,qsim_cond7,qsim_appl2} and quantum chemistry~\cite{qsim_chemistry1,qsim_chemistry2,qsim_chemistry3,qsim_chemistry4,qsim_chemistry5,qsim_chemistry6}. To simulate these quantum mechanical systems, various quantum algorithms have been proposed, including the linear combination of unitary gates, quantum signal processing, and quantum walk-based methods ~\cite{qsim_algo_LCU, qsim_algo_LCU2, qsim_algo_QSP, qsim_algo_quantumwalk, qsim_algo_quantumwalk2, qsim_algo_quantumwalk3, qsim_algo_qwalk, qsim_algo_qubitization}.

Quantum simulation is a particularly promising application as quantum computing advances beyond the NISQ era. In this early fault-tolerant regime, error-corrected logical operations become available, while logical-qubit counts and executable circuit depths remain limited~\cite{Katabarwa2024EFTQC}. At the same time, substantial classical computation and repeated circuit execution remain feasible. Together, these conditions enable hybrid coherent algorithms that go beyond the shallow variational methods developed for NISQ hardware~\cite{ZNE1,Peruzzo2014VQE}. Many such algorithms use coherent quantum simulation as a central algorithmic primitive, allowing them to calculate dynamical observables and properties of interacting quantum systems with systematically controlled algorithmic error~\cite{QZMC,LinTong2022EFTGSEE,KanSymons2025FTSimulation}.

Despite this potential, current quantum simulation algorithms face different limitations in balancing gate complexity and simulation error. Trotterization~\cite{Suzuki1,Suzuki2,Trotter_error_scaling,trotter_opt,PF_chin,PF_random,PF_random2,PF_ruth,PF_ruth2,PF_second} permits systematic error reduction by reducing the time-step size or using higher-order formulas. However, these improvements increase the gate count, which also depends explicitly on the number $L$ of Hamiltonian terms. By contrast, qDRIFT~\cite{qdrift} uses randomized circuits whose number of sampled gates is nominally independent of $L$. However, qDRIFT provides no systematic way to improve the convergence order itself. qDRIFT is only first-order method, achieving a smaller algorithmic error requires applying more randomly selected unitary gates in each circuit, thereby increasing the gate count.  These complementary limitations motivate simulation methods that retain the simple randomized circuits of qDRIFT while permitting systematic improvement of the error scaling.

In this work, we propose qSHIFT, an adaptive sampling-based protocol that resolves the trade-off between the $L$-dependent depth of Trotterization and the limited error scaling of qDRIFT. By adaptively updating the probability distribution at each sampling round, qSHIFT achieves both $L$-independence and an improved gate complexity of $\mathcal{O}_r\left((\lambda t)^{1+\frac{1}{r}}/\varepsilon^{1/r}\right)$, where $r$ is an adjustable algorithmic parameter. This gate complexity is achieved by solving a system of $L^r$ linear equations classically for each of the $N/r$ adaptive sampling rounds. Furthermore, our qSHIFT exploits the same gate set of qDRIFT unlike qSWIFT which uses additional complicated controlled unitary gates associated with ancilla qubits.

The rest of this paper is organized as follows. In Sec.~\ref{background}, we review the Trotterization and qDRIFT methods. In Sec.~\ref{protocol}, we introduce qSHIFT and analyze its algorithmic error. In Sec.~\ref{application}, we present numerical results demonstrating the performance of qSHIFT. In Sec.~\ref{discussion}, we conclude with a discussion of the results and their implications. Detailed derivations supporting the main text, along with explicit examples of the qSHIFT construction, are provided in the Appendices.

\section{Background}\label{background}
We consider the quantum simulation of a system governed by a Hamiltonian $H=\sum_{i=1}^{L} h_i H_i$, where $h_i>0$. For systems with geometrically local interactions, we can efficiently group commuting terms within $\{H_i\}$. This compression ensures the total number of grouped terms $L$ does not grow with the system size.
This form is enough to encompass a wide range of physically relevant models.

Given an initial state $|\psi\rangle$, the time-evolution of the physical observable $\mathcal{Q}$ is given by
\begin{align}
\langle\mathcal{Q} (t)\rangle_U \equiv \langle \psi | U^\dagger(t) \mathcal{Q} U(t) |\psi \rangle,
\end{align}
where $U=e^{-i H t}$. 
Since the exact implementation of $U(t)$ generally requires prohibitive quantum resources, practical simulation algorithms aim to approximate $\langle\mathcal{Q}(t) \rangle_U$ with a manageable quantum circuit. In this section, we briefly review two representative approaches: the product formula approach and the sampling-based approach.

\subsection{Product formula approach: Trotterization}
The product formula approach approximates $U(t)$ by a designed product of unitary operators. For example, the first- and second-order Trotter formulas are given by
\begin{align*}
V_2 (t) &= \prod_{i=1}^L  \left( e^{-i h_i H_i t} \right),\\
V_3 (t) &= \prod_{i=1}^L \left( e^{-i h_i H_i t/2} \right)\prod_{i=L}^1 \left( e^{-i h_i H_i t/2} \right),
\end{align*}
where the associated errors satisfy 
\begin{align*}
\langle \mathcal{Q} (t)\rangle_{V_{s}} - \langle \mathcal{Q}(t) \rangle_{U} = \mathcal{O}(t^s), \quad \text{($s$=2,3)}.
\end{align*}
More generally, other designed sequences of unitary operators with higher($s\geq3$) precision are well-known~\cite{Suzuki1,Suzuki2,Trotter_error_scaling,trotter_opt,PF_chin,PF_random,PF_random2,PF_ruth,PF_ruth2,PF_second}. 
In addition, for a given product formula, the folding of the product formula can be introduced as
\begin{align*}
V_{N,\alpha}(t) \equiv \left(V_\alpha \left(\frac{t}{N} \right)  \right)^N,
\end{align*}
which further improves the approximation accuracy since the induced algorithmic error is suppressed by an integer power of $1/N$.  

In the product formula approach, the number of gates to achieve the target precision $\varepsilon$ can be estimated analytically. For example,  the gate complexities for the first- and the second-order formula scale as $
\mathcal{O}\left(L^3 \left(\Lambda t \right)^2 /\varepsilon)\right)$, and $ \mathcal{O}\left(L^{5/2} \left(\Lambda t \right)^{3/2} /\varepsilon^{1/2})\right)$ 
respectively with $\Lambda= \text{max}_i\; h_i$~\cite{Trotter_error_scaling,qdrift}.

While product formulas provide deterministic precision without sampling overhead, they typically necessitate long gate sequences that scale with the complexity of the system.
The origin of this complexity lies in the requirement to implement every term in the Hamiltonian within each of the $N$ Trotter time steps. In systems such as complex molecules or long-range condensed matter models, the number of Hamiltonian terms can be quite large, often scaling polynomially with the number of orbitals or sites. Because the total gate count is roughly the product of $L$ and the number of steps $N$ required for precision, the resulting deep circuits become highly vulnerable to physical error accumulation. This $L$-dependent bottleneck in Trotterization provides the primary motivation for the sampling-based approaches introduced in the following subsection, which decouple circuit depth from the number of Hamiltonian terms.

\subsection{Sampling-based approach: qDRIFT}
A sampling-based approach leverages an ensemble of random circuits. A representative example of such methods is qDRIFT~\cite{qdrift}. In the $N$-qDRIFT protocol, $N(>L)$ unitary gates are sequentially sampled from an operator set 
\begin{align}
\mathcal{V} = \{ V_i(t)= e^{-i \frac{t\lambda}{N} H_i} | i=1,\cdots ,L\},
\label{operatorpool}
\end{align}
with fixed probability $p_i = h_i/\lambda$ for sampling the unitary operator $V_i(t)$ with $\lambda = \sum_i h_i$. 

After the sampling process is completed, the ensemble average of the measurement outcomes of $\mathcal{Q}$ with respect to evolved states is given by
\begin{align}
\langle \mathcal{Q}(t)\rangle_{\text{qDRIFT}}= \sum_{\vec{s}} p_{\vec{s}} \langle \psi| V^\dagger_{\vec{s}}(t)  \mathcal{Q} V_{\vec{s}}(t) | \psi \rangle,
\end{align}
where $\sum_{\vec{s}}$ is sum over all possible sequences with  
\begin{align}
\begin{aligned}
p_{\vec{s}} &= p_{s_1}\cdots p_{s_N}=\prod_{i=s_1}^{s_N} \frac{h_i }{\lambda},\\
V_{\vec{s}}(t) &= V_{s_1}(t)\cdots V_{s_N}(t),
\end{aligned}
\end{align}
where $s_{i=1\cdots N} \in \{1,\cdots, L\}$, and note that $p_{\vec{s}}$ is fixed in the qDRIFT protocol.
From the direct calculation, (see Ref .~\cite{qdrift} or Appendix \ref{QD_review}), one can show that 
\begin{align*}
\langle \mathcal{Q}(t)\rangle_U-\langle \mathcal{Q}(t)\rangle_{\text{qDRIFT}} = \mathcal{O}\left( \frac{\lambda^2 t^2}{N}\right).
\end{align*}
Since qDRIFT relies on a sampling process, we also compute the sampling complexity for a target precision $\varepsilon$, given by
\begin{align*}
N_{\text{qDRIFT,sampling}} = \mathcal{O} \left( \frac{\lambda^2t^2 }{\varepsilon^2} \right),
\end{align*}
with the details of the calculation provided in Appendix~\ref{qdvariance}.

We remark that in the qDRIFT protocol, the number of gates required to achieve a target precision $\varepsilon$ is generally independent of $L$ (except in cases where $\lambda$ scales with $L$). The $\mathcal{O}\left((\lambda t)^{2}/\varepsilon\right)$ gate complexity in qDRIFT arises because it acts as a first-order approximation of the time-evolution operator. By sampling individual terms $H_j$ according to their weights $h_j/\lambda$, the protocol ensures the ensemble average of the random circuits matches the ideal evolution only up to the first order of the Taylor expansion. Consequently, the leading-order error term scales quadratically with the evolution time $t$, effectively restricting qDRIFT to short-time simulations or necessitating a prohibitive number of samples to achieve high precision over longer durations.

\section{qSHIFT}\label{protocol}
In this section, we propose the qSHIFT protocol, a sampling-based quantum simulation protocol that achieves tunable gate complexity for given precision while sampling gates from the same set of qDRIFT.

The key idea is to replace qDRIFT's fixed sampling distribution with an adaptive one. At each round, a classical subroutine updates the distribution using the previously sampled gates. The resulting ensemble-averaged observable matches the target evolution through a higher order in $t$. We describe the protocol below.

Each run of qSHIFT generates one history-dependent trajectory (or chain). In round $p$, one sequence of $r$ gates is sampled from a distribution conditioned only on the previous samples of the same trajectory. To estimate the expectation value, we independently repeat this trajectory-generation procedure $C$ times and average the weighted measurement outcomes. Each trajectory retains its own history, which is neither shared nor pooled across repetitions.

Given input parameters $(N,r)$, assume that $r$ divides $N$, and define $M=N/r$ as the number of adaptive sampling rounds. The protocol samples one ordered sequence of $r$ gates in each round $p=1,\ldots,M$ [Fig.~\ref{scheme}]. The round-dependent distribution is determined as follows. (Further details are provided in Appendix~\ref{mechanism}.)

\subsubsection*{First round}
Throughout this section, the superscript $(p)$ labels the sampling round and does not denote a power. Specifically, $\vec{s}^{(p)}=(s_1^{(p)},\ldots,s_r^{(p)})\in\{1,\ldots,L\}^r$ denotes an ordered sequence of $r$ operator indices in round $p$, and $p_{\vec{s}^{(p)}}$ denotes its corresponding signed coefficient. Thus, $\vec{s}^{(1)}$ refers to a sequence in the first round.

In the first round,  we need an initial probability distribution for the first round sampling. To determine the probability distribution, compare two operators
\begin{equation}
\begin{aligned}
&\sum_{\vec{s}^{(1)}} p_{\vec{s}^{(1)}} V^\dagger_{\vec{s}^{(1)}}(t)  \mathcal{Q} V_{\vec{s}^{(1)}}(t),\\
={}&\sum_{\vec{s}^{(1)}}\sum_{\vec{n}} p_{\vec{s}^{(1)}} \left( \frac{i t \lambda}{N} \right)^{n_1+\cdots +n_r}\\
&\quad\times\left(\prod_{j=1}^{r}\frac{1}{n_j!}\right)
[H_{s^{(1)}_r},\cdots,[H_{s^{(1)}_1}, \mathcal{Q}]_{n_1}\cdots]_{n_r},
\end{aligned}
\label{first_sample}
\end{equation}
where $V_{\vec{s}^{(1)}}(t)$ is a sequence of $r$ operators in $\mathcal{V}$ and 
\begin{equation}
\begin{aligned}
&e^{i \frac{rt}{N}H} \mathcal{Q} e^{-i \frac{rt}{N}H},\\
=& \sum_{n=0}^{\infty} \frac{1}{n!}\left(\frac{ itr}{N} \right)^n [H,\mathcal{Q}]_n.
\end{aligned}
\label{first_target}
\end{equation}
Here, we used simplified notation for nested commutators
\begin{align*}
[A,B]_n &\equiv [\overbrace{A,[A,\cdots[A}^{\text{$n$ times}},B]]],\quad [A,B]_0\equiv B .
\end{align*}
and  
\begin{align*}
\sum_{\vec{n}} \doteq \sum_{n_1,\cdots n_r =0}^\infty. 
\end{align*}
To approximate the target operator Eq.~\ref{first_target}, match Eq.~\ref{first_sample} and Eq.~\ref{first_target} from the lowest order of $t$ sequentially. For example,  at $\mathcal{O}(t^0)$
\begin{align*}
\sum_{\vec{s}^{(1)}} p_{\vec{s}^{(1)}} \mathcal{Q} = \mathcal{Q} ,
\end{align*}
and at $\mathcal{O}(t^1)$ 
\begin{align*}
\sum_{\vec{s}^{(1)}}\sum_{i=1}^r \left(\frac{i \lambda}{N} \right)p_{\vec{s}^{(1)}} [H_{\vec{s}^{(1)}_i},\mathcal{Q}] = \sum_{j=1}^L\left(\frac{i r}{N} \right) h_j [H_j ,\mathcal{Q}],
\end{align*}
and so on. 
Requiring operator equality for arbitrary $\{H_i\}$ and $\mathcal{Q}$ imposes a separate constraint on the coefficient of each commutator $[H_j,\mathcal{Q}]$. At first order, the constraints are
\begin{align*}
\forall_j ~~\sum_{\vec{s}^{(1)}}\sum_{i=1}^r \left(\frac{i \lambda}{N} \right)p_{\vec{s}^{(1)}} \delta_{j\in \vec{s}^{(1)}_i}  =\left(\frac{i r}{N} \right) h_j.
\end{align*}

The first round contains $L^r$ unknown signed coefficients $p_{\vec{s}^{(1)}}$, one for each ordered sequence $\vec{s}^{(1)}\in\{1,\ldots,L\}^r$. We determine these coefficients by matching the Taylor expansions of Eqs.~\eqref{first_sample} and \eqref{first_target} through order $t^r$. The resulting coefficient equations initially contain redundancies because the same nested-commutator structures appear multiple times in the expansions. After collecting identical structures and removing dependent conditions, the matching procedure yields a square linear system of $L^r$ equations for the $L^r$ unknown coefficients. The detailed counting argument is given in Appendix~\ref{mechanism}.
The explicit coefficient formulas and the triangular construction used to solve these matching equations are given in Sec.~\ref{coefficient_construction}.
We remark that solving the constructed equations for $\{ p_{\vec{s}^{(1)}}\}$ can be conducted as a classical subroutine in each round of sampling without consuming quantum resources or cost.

\subsubsection*{General sampling round}
\label{general_sampling_round}

After completing the first round, consider a general round $p\in\{2,\ldots,M\}$. For a realized trajectory, the complete history
before round $p$ is
\begin{align*}
\mathcal H_{p-1}=
\left(\vec s^{(1)},\ldots,\vec s^{(p-1)}\right),
\end{align*}
and its cumulative unitary is
\begin{align*}
V_S^{(p-1)}
=
V_{\vec s^{(1)}}V_{\vec s^{(2)}}\cdots
V_{\vec s^{(p-1)}}.
\end{align*}
The coefficients in round $p$ are conditioned on this realized history,
\begin{align*}
p_{\vec s}^{(p)}
=
p_p\left(\vec s\mid\mathcal H_{p-1}\right).
\end{align*}

For a fixed history, define the conditional ensemble obtained by adding one new $r$-gate circuit sequence:
\begin{align}
\mathcal A_p(\mathcal Q\mid\mathcal H_{p-1})
=
\sum_{\vec s^{(p)}}p_{\vec s^{(p)}}\,
V_{\vec s^{(p)}}^\dagger
\left(V_S^{(p-1)}\right)^\dagger
\mathcal Q
V_S^{(p-1)}
V_{\vec s^{(p)}}.
\label{sampled}
\end{align}
The cumulative exact target after round $p$ is
\begin{align*}
\mathcal T_p(\mathcal Q)
=
e^{i(prt/N)H}\mathcal Qe^{-i(prt/N)H}.
\end{align*}
The conditional coefficients are chosen so that these two operators have identical Taylor coefficients through order $r$:
\begin{align*}
\partial_t^q
\mathcal A_p(\mathcal Q\mid\mathcal H_{p-1})
\Big|_{t=0}
=
\partial_t^q
\mathcal T_p(\mathcal Q)
\Big|_{t=0},
\quad (q=0,\ldots,r).
\end{align*}
In Sec.~\ref{coefficient_construction}, we convert these matching conditions into explicit equations for the history, target, and adaptive signed coefficients. After converting the signed coefficients into the sampling distribution described in Sec.~\ref{quasiprobability_estimator}, one sequence $\vec s^{(p)}$ is sampled and the history is updated as
\begin{align*}
V_S^{(p)} = V_S^{(p-1)}V_{\vec s^{(p)}}.
\end{align*}
The second round is recovered by setting $p=2$, and this adaptive procedure is repeated until $p=M$.

\subsubsection*{After the final round}
After all $N$ unitary operators have been sampled, we measure the observable $\mathcal{Q}$ on the initial state $|\psi\rangle$. The ensemble average of the measurement outcome is then
\begin{align*}
&\langle \mathcal{Q}(t)\rangle_{\text{qSHIFT}}\\
&=\sum_{\vec{s}^{(M)}} p_{\vec{s}^{(M)}} \langle\psi | V^\dagger_{\vec{s}^{(M)}} \left(\left(V^{(M-1)}_S\right)^\dagger \mathcal{Q} V^{(M-1)}_S \right) V_{\vec{s}^{(M)}} |\psi \rangle,
\end{align*}
and this completes our protocol.

Since each $r$-gate round matches the exact evolution through order $r$, its local truncation error is expected to be
\begin{align*}
\mathcal{O}_r\left[
\left(\frac{\lambda rt}{N}\right)^{r+1}
\right].
\end{align*}
Assuming that errors accumulate at most linearly over the $M$ rounds, the total error scales as
\begin{align}
\left|
\langle\mathcal Q(t)\rangle_U
-
\langle\mathcal Q(t)\rangle_{\mathrm{qSHIFT}}
\right|
=
\mathcal{O}_r\left(
\frac{(\lambda t)^{r+1}}{N^{r+1}}
\right),
\label{qshift_global_error}
\end{align}
which gives the gate complexity in terms of the algorithmic control parameter $r$,
\begin{align}
\mathcal{O}_r\left(
\frac{(\lambda t)^{1+1/r}}{\varepsilon^{1/r}}
\right).
\end{align}
\begin{figure}[t!]
\centering
  \includegraphics[width=1\linewidth]{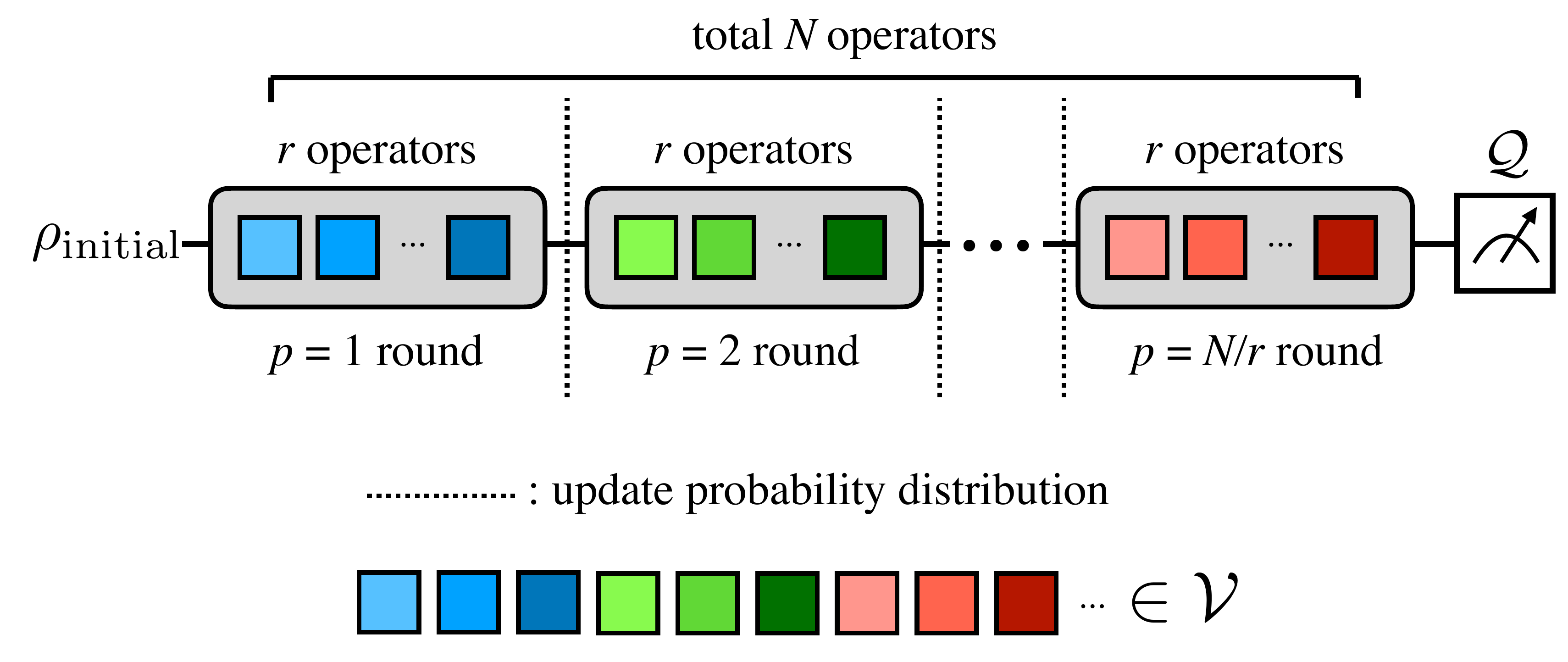}
  \caption{Schematic figure of $(N,r)$-qSHIFT. In the first round ($p=1$), sample $r$ unitary operators from $\mathcal{V}$ (Eq.~(\ref{operatorpool})) using the initial distribution. Update the distribution using the sampled operators, and then sample the next sequence. Repeat until all $N$ operators have been sampled ($p=M=N/r$). Finally, measure $\mathcal{Q}$ to complete one trajectory of $(N,r)$-qSHIFT.}
  \label{scheme}
\end{figure} 

\subsection{Calculation of the adaptive signed coefficients $p_{\vec s}^{(p)}$}
\label{coefficient_construction}

We now summarize the explicit construction of the conditional signed
coefficients in a general round. For an ordered index sequence
$\vec{a}=(a_1,\ldots,a_q)\in\{1,\ldots,L\}^q$, define
\begin{align*}
\operatorname{ad}_{\vec{a}}(\mathcal Q)
=
[H_{a_1},[H_{a_2},\ldots,[H_{a_q},\mathcal Q]\ldots]],
\end{align*}
and for the empty sequence $\varnothing$, we set $\operatorname{ad}_{\varnothing}(\mathcal Q)=\mathcal Q$. Since $V_{\vec s}=V_{s_1}\cdots V_{s_r}$, the commutator-index sequence associated with the circuit sequence $\vec s$ is
\begin{align*}
\omega(\vec s)=(s_r,\ldots,s_1).
\end{align*}
where the reversal is only an indexing convention.

For compact notation, let $x=i\lambda t/N$, and abbreviate the history and target observables as
$\mathcal Q_S^{(p-1)}=(V_S^{(p-1)})^\dagger\mathcal QV_S^{(p-1)}$ and
$\mathcal Q_T^{(p)}=e^{i(prt/N)H}\mathcal Qe^{-i(prt/N)H}$, respectively.
We define their normalized nested-commutator coefficients by the truncated expansions
\begin{align*}
\mathcal Q_S^{(p-1)}
&=
\sum_{q=0}^{r}x^q
\sum_{|\vec{ a}|=q}
S_{\vec{ a}}^{(p-1)}
\operatorname{ad}_{\vec{ a}}(\mathcal Q)
+\mathcal O(x^{r+1}),\\
\mathcal Q_T^{(p)}
&=
\sum_{q=0}^{r}x^q
\sum_{|\vec{ a}|=q}
T_{\vec{ a}}^{(p)}
\operatorname{ad}_{\vec{ a}}(\mathcal Q)
+\mathcal O(x^{r+1}).
\end{align*}
(Here, we use $|\vec{a}|$ to denote the number of entries in $\vec{a}$.)
Their empty-sequence coefficients are $S_{\varnothing}^{(p-1)}=T_{\varnothing}^{(p)}=1$. For $\vec{ a}=(a_1,\ldots,a_q)$, the target coefficient is explicit:
\begin{align}
T_{a_1\cdots a_q}^{(p)}
=
\frac{1}{q!}
\left(\frac{pr}{\lambda}\right)^q
h_{a_1}\cdots h_{a_q}.
\label{target_sequence_coefficient}
\end{align}

For two sequence-indexed coefficient arrays $A$ and $B$, define
concatenation convolution by
\begin{align}
(A*B)_{(a_1,\ldots,a_q)}
=
\sum_{k=0}^{q}
A_{(a_1,\ldots,a_k)}
B_{(a_{k+1},\ldots,a_q)},
\label{sequence_convolution}
\end{align}
where $k=0$ gives the empty prefix $\varnothing$, and $k=q$ gives the
empty suffix $\varnothing$. The sampled history uniquely determines
$S^{(p-1)}$; its explicit combinatorial formula and incremental update
are derived in Appendix~\ref{coefficient_derivation}.

Let $P_{\vec{ a}}^{(p)}$ denote the nested-commutator coefficients that must be supplied by the new round. Acting with the new block on the previously evolved observable gives the matching relation
\begin{align}
P^{(p)}*S^{(p-1)}=T^{(p)},
\label{sequence_matching}
\end{align}
through sequence length $r$, with $P_{\varnothing}^{(p)}=1$. Consequently, for every nonempty ordered index sequence $\vec{ a}=(a_1,\ldots,a_q)$, the correction coefficients are obtained successively in increasing sequence length:
\begin{align}
P_{a_1\cdots a_q}^{(p)}
={}&
T_{a_1\cdots a_q}^{(p)}
-
S_{a_1\cdots a_q}^{(p-1)}
-
\sum_{k=1}^{q-1}
P_{a_1\cdots a_k}^{(p)}
S_{a_{k+1}\cdots a_q}^{(p-1)}.
\label{sequence_recurrence}
\end{align}
This is a triangular calculation in increasing sequence length rather than solving a dense linear system. Finally, let $\mathcal B_r$ denote the
history-independent sparse linear map defined component-wise in Appendix~\ref{coefficient_derivation}, Eq.~\eqref{sequence_signed_coefficient}. It converts the coefficients $P_{\vec{ a}}^{(p)}$, indexed by ordered sequences of lengths up to $r$, into signed coefficients indexed by length-$r$ nested-commutator sequences. The coefficient assigned to the circuit sequence $\vec s$ is then
\begin{align}
p_{\vec s}^{(p)}
=
\left[\mathcal B_r P^{(p)}\right]_{\omega(\vec s)}.
\label{sol}
\end{align}
For the first round ($p=1$), there is no sampled history and $V_S^{(0)}=\mathbb{1}$. Therefore, $S_{\varnothing}^{(0)}=1$ and
$S_{\vec{ a}}^{(0)}=0$ for every nonempty sequence $\vec{ a}$. Eq.~\eqref{sequence_matching} then reduces to $P^{(1)}=T^{(1)}$, and the first-round signed coefficients are obtained directly as
\begin{align*}
p_{\vec s}^{(1)}
=
\left[\mathcal B_r T^{(1)}\right]_{\omega(\vec s)}.
\end{align*}
Here, Eq.~\eqref{target_sequence_coefficient} gives
$T_{a_1\cdots a_q}^{(1)}
=\frac{1}{q!}(r/\lambda)^q h_{a_1}\cdots h_{a_q}$.

Thus, each round consists of four classical steps:
\begin{enumerate}
\item calculate $S^{(p-1)}$ from the sampled history;
\item calculate $T^{(p)}$ from Eq.~\eqref{target_sequence_coefficient};
\item obtain $P^{(p)}$ successively from Eq.~\eqref{sequence_recurrence};
\item apply the precomputed map $\mathcal B_r$ in Eq.~\eqref{sol}.
\end{enumerate}
The explicit history formula, the convolution inverse, and the construction of $\mathcal B_r$ are provided in Appendix B2. In particular, for $r=2$, we derive a closed-form expression for every history-dependent coefficient $p_{ij}^{(p)}$, allowing all $L^2$ coefficients to be evaluated directly from the sampled history without solving a linear system at each round.

\subsubsection*{Quasi-probability sampling and complete estimator}
\label{quasiprobability_estimator}

We evaluate the linear combination in Eq. \eqref{sampled} stochastically through sampling. However, because the coefficients $\{p_{\vec{s}^{(p)}}\}$ can take negative values, they cannot be interpreted directly as standard probabilities. To address this, we adopt a quasi-probability sampling scheme to evaluate the expectation value as follows:
\begin{align}
\mathbb{E}_p(\bullet)
&=
\sum_{\vec{s}^{(p)}}p_{\vec{s}^{(p)}}\,\bullet
\nonumber\\
&=
\sum_{\vec{s}^{(p)}}q_{\vec{s}^{(p)}}Z_p
\operatorname{sign}\!\left(p_{\vec{s}^{(p)}}\right)\bullet .
\label{quasisampling}
\end{align}
Here, the normalization factor and the corresponding probability distribution in round $p$ are
\begin{align*}
Z_p
&=\sum_{\vec{s}^{(p)}}|p_{\vec{s}^{(p)}}|,
\nonumber\\
q_{\vec{s}^{(p)}}
&=\frac{|p_{\vec{s}^{(p)}}|}{Z_p}.
\end{align*}
Each execution of qSHIFT generates one trajectory by sampling one sequence of $r$ operators per round, conditioned only on the previous samples of the same trajectory. We independently repeat the complete trajectory $C$ times and average the cumulative-weighted measurement outcomes,
\begin{align}
\widehat{\mathcal Q}(t)
&=\frac{1}{C}\sum_{c=1}^{C}W_c X_c,
\label{trajectory_estimator}
\end{align}
where $X_c=\langle\psi|V_{S,c}^{\dagger}\mathcal QV_{S,c}|\psi\rangle$, and $W_c$ is accumulated round by round in Algorithm~\ref{algorithm:qSHIFT}.
The overall procedure is summarized in Algorithm \ref{algorithm:qSHIFT} and in the schematic figure [Fig.~\ref{scheme}]. 

\begin{figure}[h]
\begin{algorithm}[H]
\caption{\textbf{qSHIFT}}
\label{algorithm:qSHIFT}
\begin{algorithmic}

\STATE \textbf{Input:} $N$, $\mathcal{V}=\{V_i=e^{-i\frac{t\lambda}{N}H_i}\mid i=1,\cdots,L\}$,
$r$, initial state $|\psi\rangle$, number of independent trajectories $C$
\STATE $M\leftarrow N/r$

\FOR{$c = 1,\cdots,C$}
\STATE $V_S\leftarrow\mathbb{1}$ and $W_c\leftarrow1$

\FOR{$p = 1,\cdots,M$}

\STATE Calculate $S^{(p-1)}$ from the sampled history using Eq.~\eqref{history_series}
\STATE Calculate $T^{(p)}$ using Eq.~\eqref{target_sequence_coefficient}
\STATE Calculate $P^{(p)}$ successively using Eq.~\eqref{sequence_recurrence}
\STATE Obtain $p_{\vec{s}^{(p)}}$ from the sparse map in Eq.~\eqref{sol}
\STATE Set $Z_p=\sum_{\vec{s}}|p_{\vec{s}^{(p)}}|$ and $q_{\vec{s}^{(p)}}=|p_{\vec{s}^{(p)}}|/Z_p$
\STATE Sample one set of $r$ unitary operators $V_{\vec{s}}$ with probability $q_{\vec{s}^{(p)}}$
\STATE $W_c\leftarrow W_c Z_p\,\mathrm{sign}(p_{\vec{s}^{(p)}})$ for the sampled $\vec{s}$
\STATE $V_S \leftarrow V_S V_{\vec{s}}$
\ENDFOR

\STATE Measure $X_c=\langle \psi | V_S^\dagger \mathcal{Q} V_S | \psi \rangle$
\ENDFOR

\STATE \textbf{Output:}
\begin{align*}
\widehat{\mathcal Q}(t)=\frac{1}{C}\sum_{c=1}^{C}W_cX_c.
\end{align*}
\end{algorithmic}
\end{algorithm}
\end{figure}

We conclude this section with two remarks. First, $(N,r=1)$-qSHIFT without updating probability distribution, i.e. fixed probability distribution converges to $N$-qDRIFT. The other remark is that the protocol can be easily generalized by replacing the fixed drawing number $r$ with arbitrary integers $\{ s_i \}$ satisfying $\sum_i s_i = N$.

\section{Numerical demonstration}\label{application}

\subsection{Quantum simulation}
In this section, we apply our $(N,r)$-qSHIFT protocol to 1D transverse field Ising model,
\begin{align*}
H&= h_1 H_1 + h_2 H_2,\\
H_1&=\sum_{i=1}^{7} Z_iZ_{i+1},\\
H_2&=\sum_{i=1}^{8} X_i,
\end{align*}
with $(h_1=J,h_2=h)= (1, 0.1)$ on a system of eight qubits.
This model serves as a standard benchmark for quantum simulation algorithms.
For a randomly chosen initial state, we estimate the time-evolution of a physical observable $\mathcal{Q} =\sum_{i=1}^{8} Z_i$.

We perform simulation using $(N=2,r=2)$-qSHIFT and $(N=3,r=3)$-qSHIFT.  For the $(N=2,r=2)$-qSHIFT, the probability distribution obtained from our protocol is 
\begin{align}
\begin{aligned}
p_{11} &= \frac{h_1 }{\lambda} \frac{h_1-h_2 }{\lambda} ,\\
p_{12}&=p_{21}= 2\frac{h_1 }{\lambda}\frac{h_2 }{\lambda},\\
p_{22} &= \frac{h_2 }{\lambda} \frac{h_2-h_1 }{\lambda}.
\end{aligned}
\end{align}
Note that since $h_1>h_2$, we have $p_{22}<0$, and the quasi-probability sampling scheme in Sec~\ref{protocol} is applied. 
For $(N=3,r=3)$-qSHIFT, the corresponding probabilities are
\begin{align}
\begin{aligned}
p_{iii} &= \frac{1}{6} \left(\frac{3h_i}{\lambda}\right)^3-\frac{1}{2} \left(\frac{3h_i}{\lambda}\right)^2  + \frac{1}{3} \left(\frac{3h_i}{\lambda}\right),\\
p_{ijj} &= \frac{1}{6} \left(\frac{3h_i}{\lambda}\right)\left(\frac{3h_j}{\lambda}\right)^2 -\frac{1}{4} \left(\frac{3h_i}{\lambda}\right)\left(\frac{3h_j}{\lambda}\right), \\
p_{iij}&= \frac{1}{6} \left(\frac{3h_i}{\lambda}\right)^2 \left(\frac{3h_j}{\lambda}\right) - \frac{1}{4} \left(\frac{3h_i}{\lambda}\right)\left(\frac{3h_j}{\lambda}\right),\\
p_{iji}&= \frac{1}{6} \left(\frac{3h_i}{\lambda}\right)^2\left(\frac{3h_j}{\lambda}\right),
\end{aligned}
\end{align}
with $i,j\in \{1,2\}$ and $i\neq j$. 
Detailed calculations for these probability distributions are provided in Appendix~\ref{example}. 
\begin{figure}[t!]
\centering
  \includegraphics[width=1\linewidth]{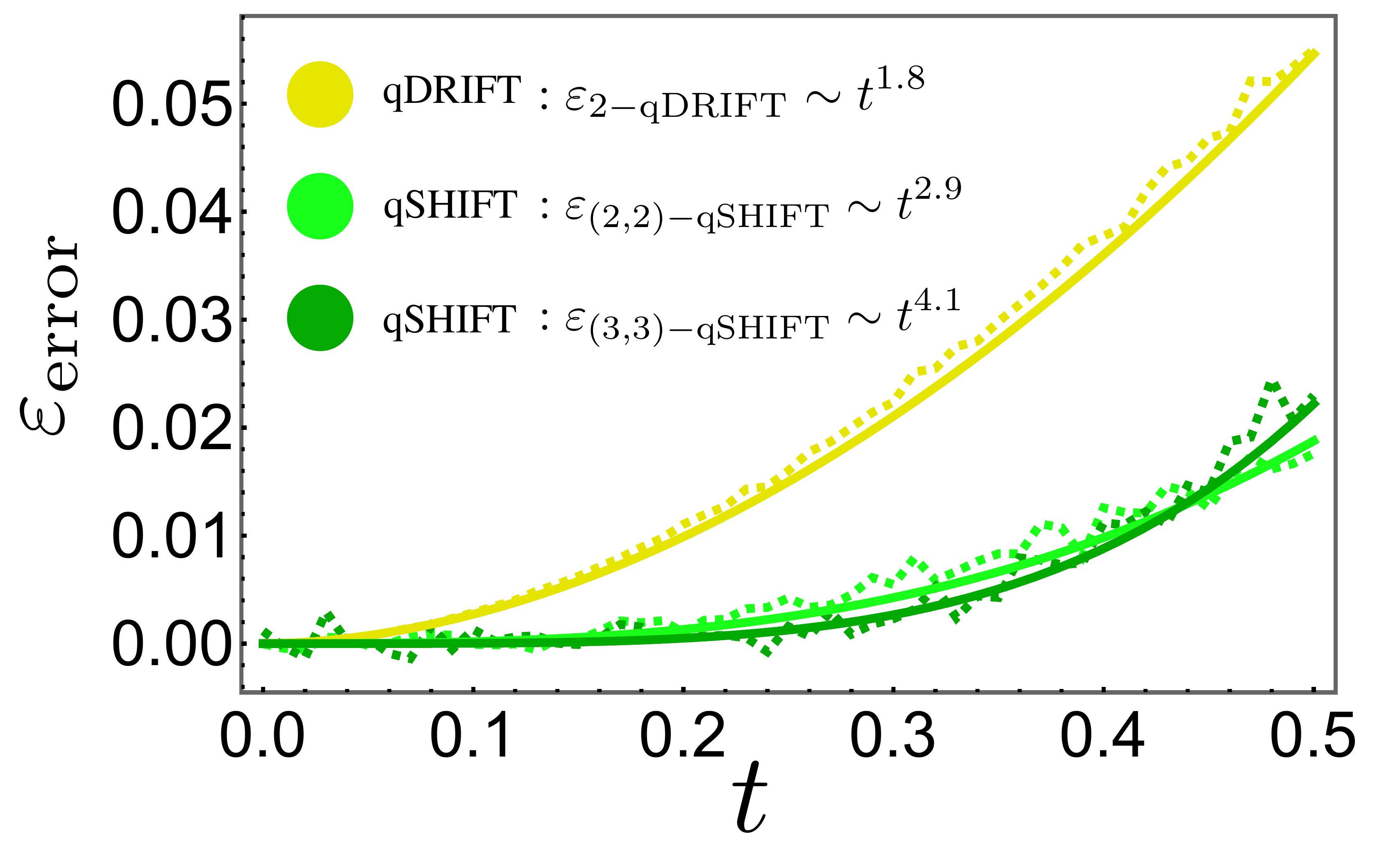}
  \caption{Numerical demonstrations the $(N,r)$-qSHIFT and qDRIFT protocols for the 1D transverse field Ising model with $(h_1,h_2)=(1,0.1)$ on eight qubits. The time-evolution of a physical observable $\mathcal{Q}= \sum_{i=1}^{8} Z_i$ is estimated for a randomly chosen initial state. Absolute algorithmic errors $\varepsilon_{\text{error}}= |\langle \mathcal{Q} (t) \rangle _{\text{sampled}}- \langle \mathcal{Q} (t) \rangle _{\text{exact}}|$,  mitigated by the qSHIFT protocol (green dashed lines) and the qDRIFT protocol (yellow dashed lines) are shown as a function of evolution time $t$. Power-law fitting of mitigated errors confirms that the theoretically expected scalings: $\mathcal{O}(t^2)$ for qDRIFT and $\mathcal{O}(t^{1+r})$ for $(N,r)$-qSHIFT.}
  \label{numerics}
\end{figure} 
In Fig.~\ref{numerics}, we compare the algorithmic errors of the different protocols, including $(N=2)$-qDRIFT as a benchmark. 
Power-law fitting of the mitigated errors confirms the theoretically predicted scaling behaviors, yielding $\mathcal{O}(t^{1+r})$ for $(N,r)$-qSHIFT, and $\mathcal{O}(t^{2})$ for qDRIFT, demonstrating that qSHIFT achieves a systematic improvement in error scaling controlled by the parameter $r$.

\begin{table*}[t!]
\centering
\begin{tabular}{|c|c|c|}
\hline
\textbf{Protocol} & \textbf{Gate complexity} & \textbf{Sampling complexity}\\
\hline
$1^{\text{st}}$ order Trotter formula
& $\mathcal{O}\left(L^{3}(\Lambda t)^{2}/\varepsilon\right)$ & \text{(Deterministic)} \\
\hline
$2^{\text{nd}}$ order Trotter formula
& $\mathcal{O}\left(L^{5/2}(\Lambda t)^{3/2}/\varepsilon^{1/2}\right)$ &\text{(Deterministic)}\\
\hline
$(2k)^{\text{th}}$ order Trotter formula
& $\mathcal{O}\left(L^{2+\frac{1}{2k}}(\Lambda t)^{1+\frac{1}{2k}}/\varepsilon^{1/2k}\right)$& \text{(Deterministic)} \\
\hline
$N$-qDRIFT (general result)
& $\mathcal{O}\left((\lambda t)^{2}/\varepsilon\right)$ & $\mathcal{O} \left( (\lambda t)^2/  \varepsilon^2\right) $ \\
\hline
$K^{\text{th}}$ order qSWIFT
& $\mathcal{O}\left(\left( \lambda t\right)^2 /\varepsilon^{1/K}\right)$ & $\mathcal{O}\left( 
\frac{K+1}{\varepsilon^2} \sum_{k=1}^K  \left(\frac{(2\lambda t)^2 }{N} \right)^k \right)$ \\
\hline
{$(N,r)$-qSHIFT (this work)}
& $\mathcal{O}\left((\lambda t)^{1+\frac{1}{r}}/\varepsilon^{1/r}\right)$ & \makecell[l]{
$\mathcal{O}\left( \left( \frac{ (M!)^{2r} e^{\Theta(M)} \langle\mathcal{Q} \rangle^2 } {\varepsilon^2} \right) \right)$, \text{for $Z(p_{\vec{s}})>1$,}\\
$\mathcal{O}\left( \frac{\lambda^2 t^2 }{\varepsilon^2} \right)$,  ~~~\quad\quad\quad~\;\text{for $Z(p_{\vec{s}})=1$.} }\\
\hline
\end{tabular}
\caption{ The gate complexity and sampling complexity required to achieve a target precision $\varepsilon$ of an observable for several Hamiltonian simulation protocols. 
The results of gate complexity for qDRIFT, qSWIFT and the standard Trotter formulas are adapted from Ref.~\cite{qdrift,qswift}, while the $(N,r)$-qSHIFT entries correspond to the present work. 
Scalings are expressed in terms of the number of Hamiltonian terms $L$, the evolution time $t$, and the coefficient scales $\Lambda=\text{max}_i h_i$ and $\lambda=\sum_i h_i$. 
While the gate complexity of deterministic Trotter formula approaches generally scales with $L$ and it is deterministic protocol, the gate complexities of sampling-based protocols, both qDRIFT and qSHIFT, are independent of $L$. We plot asymptotic behaviors of gate complexities in Fig.~\ref{gate_count}.  
The sampling complexities of sampling based protocols are presented in the third column.
For $K^{\text{th}}$ order qSWIFT,  $K$ is an algorithmic parameter of qSWIFT~\cite{qswift}.    }
\label{gatenumber}
\end{table*}

\section{Discussion and Conclusion}\label{discussion}
\begin{figure*}[t!h!]
\centering
  \includegraphics[width=1\textwidth]{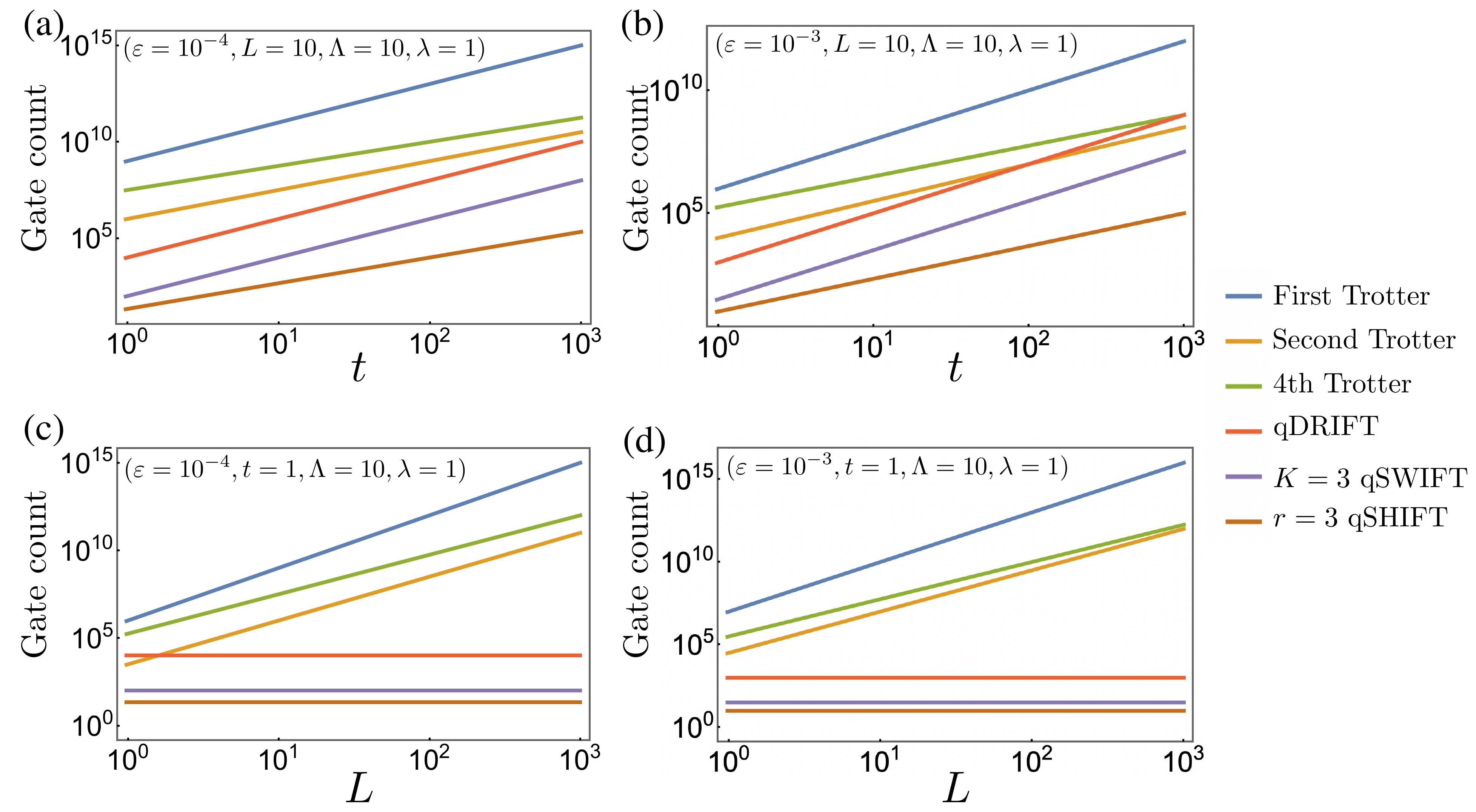}
  \caption{Gate counts scalings for the fixed precision $\varepsilon$.  We plot scaling behaviors of gate complexities for various algorithms. Depending on various parameters,  we observe the scaling behaviors of gate complexities. The brown lines represent estimated gate complexities of $r=3$-qSHIFT. Panel (a), (b) show gate complexity with respect to $t$ and panel (c) and (d) show gate complexity with respect to the number of Hamiltonian components $L$. }
  \label{gate_count}
\end{figure*}

In this work, we proposed qSHIFT, a sampling-based quantum simulation protocol that adaptively updates the probability distributions for circuit sampling at each round. By sampling $r$ operators in each round, qSHIFT achieves a gate complexity $\mathcal{O}_r\left((\lambda t)^{1+1/r}/\varepsilon^{1/r}\right)$ which is $L$-independent as qDRIFT. We summarize and compare qSHIFT with existing methods in Table~\ref{gatenumber}. As shown, $(N,r)$-qSHIFT achieves improved scaling in both $t$ and $\varepsilon$ compared to qDRIFT and qSWIFT. The results for qDRIFT and standard Trotter formulas are adapted from Ref.~\cite{qdrift,qswift}, while the $(N,r)$-qSHIFT results are derived in this work. We also plot a gate complexity behaviors for various parameters, see Fig.~\ref{gate_count}.
Consequently, qSHIFT provides a practical balance between accuracy and implementability on early fault-tolerant quantum devices.
This improvement is obtained at the classical cost of solving a system of $L^r$ linear equations in a classical subroutine for each of the $M$ sampling rounds. Since these adaptive updates can be computed on a classical computer, the protocol allows for the flexible pre-configuration of quantum circuits before execution without introducing additional quantum overhead.

We also analyzed the sampling complexity of the qSHIFT protocol for a target precision $\varepsilon$. The complexity $N_{\text{qSHIFT,sampling}}$ is given by:
\begin{align*}
N_{\text{qSHIFT,sampling}} = 
\begin{cases}
\mathcal{O}\left( \frac{(M!)^{2r}e^{\Theta(M)}} {\varepsilon^2} \langle\mathcal{Q} \rangle^2\right), &\text{for } Z(p_{\vec{s}})>1,\\
\mathcal{O}\left( \frac{\lambda^2 t^2 }{\varepsilon^2}  \right), &\text{for } Z(p_{\vec{s}})=1.
\end{cases}
\end{align*}
where $M=N/r$ is the number of total adaptive sampling round. 
When all $p_{\vec{s}} > 0$, $Z(p_{\vec{s}}) = 1$ and the sampling complexity depends only on $\lambda$ and $t$, matching qDRIFT. In contrast, when $Z(p_{\vec{s}}) > 1$, the sampling overhead is proportional to $\langle \mathcal{Q}\rangle^2$. 
Thus, sampling with a single round will reduce sampling overhead. Detailed calculations are provided in Appendix~\ref{variance}. Furthermore, the sampling complexity of qSHIFT scales polynomially with the number of qubits $N_{\text{sys}}$: in the $Z(p_{\vec{s}})=1$ case, it is governed by $\lambda^2 \sim \mathcal{O}(N_{\text{sys}}^2)$, while in the $Z(p_{\vec{s}})>1$ case, it is proportional to $\langle \mathcal{Q}\rangle^2$, which is also typically $\mathcal{O}(N_{\text{sys}}^2)$ for local observables.

Let us remark on existing sampling based quantum simulation protocols, qDRIFT\cite{qdrift} and qSWIFT\cite{qswift} compared to our qSHIFT protocol. First, qSHIFT exploits the same gate set (Eq.~\ref{operatorpool})
of qDRIFT but the difference is that qSHIFT leverages classical subroutine for adaptive probability update for sampling each round. Such classical subroutine which does not cost any quantum resources and deteriorate the outcome and make qSHIFT achieve faster algorithm compared to qDRIFT. Secondly, qSWIFT leverages an ancilla to realize the algorithm and controlled gates to realize one of main channel: swift channel. To realize such channels requires a number of entangling gates, which results in deepening native circuit depth whereas our qSHIFT uses the same gate set of qDRIFT without the ancilla qubit and associated controlled gates. Thus, qSHIFT is well suited to early fault-tolerant quantum devices with limited logical qubits and executable circuit depth.

Notably, in the $N=r\to \infty$ limit, qSHIFT approaches the optimal complexity $\mathcal{O}(t)$ implied by the no fast-forwarding theorem~\cite{nfft1, nfft2}, indicating that the adaptive sampling strategy is asymptotically optimal. These features establish qSHIFT as a robust framework for accurate and resource-efficient quantum simulation.

Compared with conventional algorithms such as Trotterization and quantum signal processing, qSHIFT requires fewer quantum gates, enhancing its robustness to physical noise. This makes qSHIFT particularly compatible with error mitigation techniques like zero-noise extrapolation and probabilistic error cancellation~\cite{PEC1,PEC2,ZNE1,ZNE2}. Since the overhead of these methods grows rapidly with circuit depth, the reduced depth of qSHIFT can significantly lower the total mitigation cost. Combining qSHIFT with such strategies represents a promising avenue for improving practical quantum simulation performance.

Several directions remain for future investigation. First, since qSHIFT can be integrated as a subroutine into broader algorithms like qSWIFT~\cite{qswift} and Krylov-based diagonalization~\cite{sqdrift}, exploring the performance gains in these contexts is a natural next step. Second, developing more efficient classical algorithms for generating the adaptive distributions would further improve practicality. Third, incorporating symmetry constraints into the sampling process could preserve selection rules and reduce sampling complexity. Finally, benchmarking qSHIFT on real quantum hardware will be essential for evaluating its practical utility across the transition from late-NISQ to early fault-tolerant quantum computing.

\section{Acknowledgment}
The authors thank Hyukjoon Kwon and  Hyukgun Kwon for helpful comments. SC was supported by a KIAS Individual Grant (CG090601) at Korea Institute for Advanced Study and by Quantum Simulator Development Project for Materials Innovation through the National Research Foundation of Korea (NRF) funded by the Korean government (Ministry of Science and ICT(MSIT))(No. NRF- 2023M3K5A1094813).  SL is supported by a KIAS Individual Grant via the Quantum Universe Center (QP104301-6P104301) at Korea Institute for Advanced Study and by Institute of Information \& communications Technology Planning \& Evaluation (IITP) grant funded by the Korea government (MSIT) (No.2022-0-01026).  We used resources of the Center for Advanced Computation at Korea Institute for Advanced Study.

 \bibliography{qshiftref}

\appendix
\clearpage
\onecolumngrid
\section{Details of qDRIFT calculations}
\label{QD_review}
\subsection{Mechanism of the qDRIFT}
In this section, we review the original qDRIFT algorithms. 
We consider a Hamiltonian,
\begin{align*}
H= \sum_{i=1}^{L} h_i H_i, ~(\forall_i  h_i>0),
\end{align*}
and a physical observable $\mathcal{Q}$ that we want to simulate with respect to an initial state $|\psi\rangle$. 
The purpose of quantum simulation is to estimate
\begin{align*}
\langle\mathcal{Q} (t)\rangle_{\text{ideal}} = \langle\psi | e^{i t H} \mathcal{Q} e^{-i t H}|\psi \rangle,
\end{align*}
as precisely as possible using quantum circuits.

To esmitate $\langle\mathcal{Q} (t)\rangle_{\text{ideal}}$ with qDIRFT, consider a set of operators, 
\begin{align*}
\mathcal{V}=\big\{ V_{i=1,\cdots,L}\big| V_i=e^{-i \frac{\lambda t}{N} H_i } \big\},
\end{align*}
where $\lambda= \sum_i h_i$ and $N(>L)$ is an integer parameter of qDRIFT. 
The qDRIFT uses a probability distribution 
\begin{align*}
\mathcal{P} = \Big \{p_{i=1,\cdots,L}\big | p_i=\frac{h_i}{\lambda} \Big\},
\end{align*}
of drawing an operator $V_i$ in $\mathcal{V}$. The qDRIFT simulate the quantum system with the generated quantum circuit with drawn $N$ unitary operators. 

After drawing of $N$ operators from $\mathcal{V}$ following the probability distribution $\mathcal{P}$, the expectation value of $\mathcal{Q}$ is
\begin{align*}
\langle\mathcal{Q} (t)\rangle_{\text{qDRIFT}} &= p_1^N \langle V^{\dagger N}_1 \mathcal{Q} V^N_1  + \cdots + p_L^N \langle V^{\dagger N}_L \mathcal{Q} V^N_L\rangle,\\
&= \sum_{s_1,\cdots,s_N=1}^{L} p_{s_1} p_{s_2} \cdots p_{s_N} \langle V^\dagger_{s_1} V^\dagger_{s_2} \cdots V^\dagger_{s_N}\mathcal{Q} V_{s_1} V_{s_2} \cdots V_{s_n}\rangle,\\
&=\sum_{\vec{s}}  p_{\vec{s}} \langle V^\dagger_{\vec{s}} \mathcal{Q} V_{\vec{s}}\rangle,\\
&=\sum_{n_1,\cdots,n_N=0}^\infty \sum_{\vec{s}} p_{\vec{s}} \left( \frac{i t \lambda}{N} \right)^{n_1+\cdots n_N}\langle [H_{s_1},[H_{s_2},\cdots,[H_{s_N},\mathcal{Q}]_{n_N}]_{n_{N-1}}\cdots ]_{n_2}]_{n_1}\rangle,\\
&= \left(\sum_{\vec{s}} p_{\vec{s}} \langle \mathcal{Q}\rangle \right) + \left(\frac{i t \lambda}{N}  \right)\sum_{\vec{s}} p_{\vec{s}}  \langle \left(  [H_{s_1},\mathcal{Q}] +\cdots  [H_{s_N},\mathcal{Q}] \right)\rangle+ \mathcal{O}\left(\left(\frac{t\lambda}{N}\right)^2 \right),\\
&= \langle \mathcal{Q} \rangle+ \left(\frac{i t \lambda}{N}\right) N  \sum_{i=1}^L  p_{s_i} \langle [H_{s_i},\mathcal{Q}] \rangle +\mathcal{O}\left(\left(\frac{t\lambda}{N}\right)^2 \right),\\
&= \langle \mathcal{Q} \rangle + \left(i t \lambda \right) \sum_{i=1}^L  \frac{h_{s_i}}{\lambda}  \langle [H_{s_i},\mathcal{Q}] \rangle +\mathcal{O}\left(\left(\frac{t\lambda}{N}\right)^2 \right),\\
&= \langle \mathcal{Q}\rangle + i t \langle [H,\mathcal{Q}]\rangle + \mathcal{O}\left(\left(\frac{t\lambda}{N}\right)^2 \right).
\end{align*}
In the second line, we defined
\begin{align*}
p_{\vec{s}} = p_{s_1} \cdots p_{s_N},~ \text{and}~V_{\vec{s}}= V_{s_1} \cdots V_{s_N},
\end{align*}
for brevity of notations, and also introduced summation
\begin{align*}
\sum_{s_1,\cdots s_N=1}^{L} \equiv \sum_{\vec{s}} ,
\end{align*}
meaning that the sum over all possible $s_{i=1\cdots N}$. 
In the third line, we used
\begin{align*}
e^{ A} B e^{- A} &= \sum_{n=0}^\infty \frac{1}{n!}[A,B]_n,
\end{align*}
where
\begin{align*}
[A,B]_n &\equiv [\overbrace{A,[A,\cdots[A}^{\text{$n$ times}},B]]],\quad [A,B]_0\equiv B .
\end{align*}
Since the ideally time evolved expectation value is
\begin{align*}
\langle \mathcal{Q}(t) \rangle _{\text{ideal}} 
&=\sum_n \frac{1}{n!} \left( i t \right)^n \langle[H,\mathcal{Q}]_n\rangle,\\
&=\langle \mathcal{Q}\rangle + \left( i t \right) \langle [H,\mathcal{Q}]\rangle + \frac{1}{2} \left( i t \right)^2 \langle [H,[H,\mathcal{Q}]]\rangle + \cdots,
\end{align*}
and thus an algorithmic error of the qDRIFT protocol is $\mathcal{O}(t^2)$.

\subsection{Variance of qDRIFT protocol}
\label{qdvariance}
In this section, we compute the variance of the qDRIFT protocol to estimate sampling complexity.  The variance of qDRIFT protocol is defined as
\begin{align*}
&\text{var}(\langle\mathcal{Q} \rangle_{qDRIFT})=   \sum_{s_1,\cdots s_N=1}^L p_{\vec{s}} \langle \psi | V^\dagger _{\vec{s} } \mathcal{Q} V_{\vec{s} } |\psi \rangle ^2 -  \left(\sum_{s_1,\cdots s_N=1}^L p_{\vec{s}} \langle \psi | V^\dagger _{\vec{s} } \mathcal{Q} V_{\vec{s} } |\psi \rangle \right)^2 
\end{align*}
where $p_{\vec{s}} = p_{s_1}\cdots p_{s_N}$ and $V_{\vec{s}} = V_{s_1} \cdots V_{s_N}$. 

The first term is
\begin{align*}
&\sum_{s_1,\cdots s_N=1}^L p_{\vec{s}} \langle \psi | V^\dagger _{\vec{s} } \mathcal{Q} V_{\vec{s} } |\psi \rangle ^2\\
=&\sum_{s_1,\cdots s_N=1}^L p_{\vec{s}} \left( \langle \mathcal{Q}\rangle   + \left( \frac{i\lambda t}{N} \right)\sum_{i} \langle [H_{s_i},\mathcal{Q}] \rangle  + \left(\frac{i t \lambda}{N} \right)^2 \sum_{i,j} \sum_{\substack{n_0,n_1=0,\\n_0+n_1=2}}^2  \frac{1}{n_0 ! n_1 !} \langle[H_{s_i},[H_{s_j},\mathcal{Q}]_{n_0} ]_{n_1}\rangle  + \mathcal{O}(t^3)\right)^2\\
=&\sum_{s_1,\cdots s_N=1}^L p_{\vec{s}} \langle \mathcal{Q}\rangle^2 +  2\left(\frac{i t \lambda}{N} \right) \langle \mathcal{Q} \rangle \sum_{s_1,\cdots s_N=1}^L p_{\vec{s}} \sum_{i} \langle [H_{s_i},\mathcal{Q}]\rangle \\
&\quad + \left(\frac{i t \lambda}{N} \right)^2 \sum_{s_1,\cdots s_N=1}^L p_{\vec{s}} \left( 2 \langle \mathcal{Q} \rangle \sum_{i,j} \sum_{\substack{n_0,n_1=0,\\n_0+n_1=2}}^2  \frac{1}{n_0 ! n_1 !} \langle[H_{s_i},[H_{s_j},\mathcal{Q}]_{n_0} ]_{n_1} \rangle  +  \sum_{i,j}\langle [H_{s_i},\mathcal{Q}] \rangle \langle [H_{s_j},\mathcal{Q}] \rangle   \right) +\mathcal{O}(t^3)
\end{align*}

The second term is 
\begin{align*}
  &\left(\sum_{s_1,\cdots s_N=1}^L p_{\vec{s}} \langle \psi | V^\dagger _{\vec{s} } \mathcal{Q} V_{\vec{s} } |\psi \rangle \right)^2\\
  =&\left(\sum_{s_1,\cdots s_N=1}^L p_{\vec{s}} \langle \mathcal{Q} \rangle + \left( \frac{i t \lambda}{N}\right)\sum_{s_1,\cdots s_N=1}^L p_{\vec{s}} \sum_{i}   \langle [H_{s_i},\mathcal{Q}]\rangle   + \left(\frac{i t \lambda}{N} \right)^2 \sum_{s_1,\cdots s_N=1}^L p_{\vec{s}}  \sum_{i,j} \sum_{\substack{n_0,n_1=0,\\n_0+n_1=2}}^2  \frac{1}{n_0 ! n_1 !} \langle[H_{s_i},[H_{s_j},\mathcal{Q}]_{n_0} ]_{n_1} \rangle +\mathcal{O}(t^3)\right)^2\\
  =& \langle \mathcal{Q}\rangle  ^2  +  2 \left( \frac{i t \lambda}{N} \right)\langle  \mathcal{Q} \rangle \sum_{s_1,\cdots s_N=1}^L p_{\vec{s}} \sum_{i}   \langle [H_{s_i},\mathcal{Q}]\rangle \\
&  \quad +\left(\frac{i t \lambda}{N} \right)^2 \left(2 \sum_{s_1,\cdots s_N=1}^L p_{\vec{s}} \langle \mathcal{Q} \rangle \sum_{s_1,\cdots s_N=1}^L p_{\vec{s}}  \sum_{i,j} \sum_{\substack{n_0,n_1=0,\\n_0+n_1=2}}^2  \frac{1}{n_0 ! n_1 !} \langle[H_{s_i},[H_{s_j},\mathcal{Q}]_{n_0} ]_{n_1}\rangle   + \sum_{\substack{s_1,\cdots s_N=1,\\s'_1,\cdots s'_N=1}}^L p_{\vec{s}} p_{\vec{r'}} \sum_{i,j}   \langle [H_{s_i},\mathcal{Q}]\rangle   \langle [H_{r'_i},\mathcal{Q}]\rangle\right) \\
&\quad +\mathcal{O}(t^3)
\end{align*}
We used $\sum_{s_1,\cdots s_N=1}^L p_{\vec{s}} =1$. 

Thus,  the variance of outcomes from the qDRIFT protocol is
\begin{align*}
&\text{var}(\langle\mathcal{Q} \rangle_{qDRIFT})   \\
=& \left(\frac{i t \lambda}{N} \right)^2\sum_{s_1,\cdots s_N=1}^L p_{\vec{s}} \sum_{i,j} \langle [H_{s_i},\mathcal{Q}]\rangle \left(\langle [H_{s_j},\mathcal{Q}]\rangle -\sum_{s'_1,\cdots s'_N=1}^L p_{\vec{s'}} \langle [H_{s'_j },\mathcal{Q}]\rangle   \right) + \mathcal{O}(t^3) .
\end{align*}
Hence, the sampling complexity of the qDRIFT $N_{\text{qDRIFT,sample}}$ for precision  $\varepsilon_{\text{qDRIFT}} =\mathcal{O}\left(\frac{t\lambda}{N} \right)<1$ is 
\begin{align}
N_{\text{qDRIFT,sample}} &= \frac{\text{var}(\langle\mathcal{Q} \rangle_{qDRIFT})}{\varepsilon_{\text{qDRIFT}}^2}=\mathcal{O}\left( \left(t \lambda \right)^2/\varepsilon_{\text{qDRIFT}}^2 \right).
\end{align}

\section{qSHIFT -  Formalism}
\label{App-qshift}
\subsection{Mechanism of algorithm}
\label{mechanism}
In this section, we provide detailed calculations for the main classical subroutine in qSHIFT.
To determine $p_{\vec{s}}$ for sampling, $\sum_{\vec{s}} p_{\vec{s}} V^\dagger_{\vec{s}}  \mathcal{Q} V_{\vec{s}}$ to $e^{i t \frac{ r}{N} H} \mathcal{Q}e^{-i t\frac{ r}{N} H}$ from lowest order. 

We expand $\sum_{\vec{s}} p_{\vec{s}} V^\dagger_{\vec{s}} \mathcal{Q} V_{\vec{s}}$,
\begin{align*}
&\sum_{\vec{s}}p_{\vec{s}} V^\dagger_{\vec{s}}  \mathcal{Q} V_{\vec{s}}\\
=&\sum_{\vec{s}}\sum_{n_1,\cdots,n_{r}=0}^\infty  p_{\vec{s}} \left( \frac{i t \lambda}{N} \right)^{n_1+\cdots n_{r}} [H_{s_r},\cdots[H_{s_3},[H_{s_{2}},[H_{s_1},\mathcal{Q}]_{n_{1}}]_{n_{1}}\cdots ]_{n_r},\\
=& \sum_{\vec{s}} p_{\vec{s}} \mathcal{Q}\\
& +  \sum_{\vec{s}} p_{\vec{s}} \left( \frac{i \lambda  t}{N}\right)  [H_{s_\mu},\mathcal{Q}]  \\
& + \sum_{\substack{n_1,n_2=0,\\n_1+n_1=2}}^2  \sum_{\vec{s}} p_{\vec{s}} \left( \frac{i \lambda  t}{N}\right)^2   \frac{1}{n_1 ! n_2 !}[H_{s_i},[H_{s_\mu},\mathcal{Q}]_{n_1}]_{n_2}\\
& + \sum_{\substack{n_1,n_2=0,\\n_1+n_2+n_3=3}}^3  \sum_{\vec{s}} p_{\vec{s}} \left( \frac{i \lambda  t}{N}\right)^3   \frac{1}{n_1 ! n_2 !n_3!}[H_{s_j},[H_{s_i},[H_{s_\mu},\mathcal{Q}]_{n_1}]_{n_2}]_{n_3}\\
& ~\vdots\\
& + \sum_{\substack{n_1,\cdots n_r =0,\\ \sum_{i=0}^r n_i =r}}^{r} \sum_{\vec{s}} p_{\vec{s}} \left( \frac{i \lambda  t}{N}\right)^{r}  \frac{1}{n_1! \cdots n_r!} [H_{s_r},\cdots[H_{s_3},[H_{s_{2}},[H_{s_1},\mathcal{Q}]_{n_{1}}]_{n_{2}}\cdots ]_{n_r}, \\
&+\mathcal{O}(t^{r+1})
\end{align*}
and $e^{i t \frac{ r}{N} H} \mathcal{Q}e^{-i t \frac{r}{N} H}$
\begin{align*}
&e^{i t \frac{ r}{N} H} \mathcal{Q}e^{-i t \frac{ r}{N}H},\\
=& \sum_{n} \frac{1}{n!}  \left( \frac{i t  r}{N} \right)^n [H,\mathcal{Q}]_n,\\
=& \mathcal{Q}\\
& +\sum_{i=1}^L \left(\frac{i t r}{N} \right)  [h_iH_i ,\mathcal{Q}]\\
&+ \sum_{i,j=1}^L \frac{1}{2}\left(\frac{i t r}{N}\right)^2 [h_i H_i ,[h_j H_j,\mathcal{Q}]] \\
&+ \sum_{i,j,k=1}^L \frac{1}{3!} \left(\frac{i t r}{N}\right)^3 [h_k H_k, [h_i H_i ,[h_j H_j,\mathcal{Q}]] \\
&~\vdots \nonumber\\
& + \sum_{s_1,\cdots,s_r =1}^L \frac{1}{r!}\left(\frac{i t r}{N}\right)^{r} [h_{s_r} H_{s_r},\cdots[h_{s_1} H_{s_1},\mathcal{Q}]]\cdots ]\\
&+\mathcal{O}(t^{r+1}).
\end{align*}

Comparing two operators order by order gives
\begin{align*}
\mathcal{O}(t^0)&: \sum_{\vec{s}} p_{\vec{s}} \mathcal{Q} = \mathcal{Q} ,\\
\mathcal{O}(t^1)&:   \sum_{\vec{s}} p_{\vec{s}} \left( \frac{i \lambda  t}{N}\right)  [H_{s_\mu},\mathcal{Q}]  =\sum_{i=0}^L \left(\frac{i t r }{N} \right)  h_i[H_i ,\mathcal{Q}],\\
\mathcal{O}(t^2)&:  \sum_{\substack{n_0,n_1=0,\\n_0+n_1=2}}^2  \sum_{\vec{s}} p_{\vec{s}} \left( \frac{i \lambda  t}{N}\right)^2   \frac{1}{n_0 ! n_1 !}[H_{s_i},[H_{s_\mu},\mathcal{Q}]_{n_0}]_{n_1}
=\sum_{i,j=1}^L \frac{1}{2}\left(\frac{i t r}{N}\right)^2 [h_i H_i ,[h_j H_j,\mathcal{Q}]], \\
&\vdots\nonumber\\
\mathcal{O}(t^{r} )&: \sum_{\substack{n_0,\cdots n_r =0,\\ \sum_{i=0}^r n_i =r }}^{r} \sum_{\vec{s}} p_{\vec{s}} \left( \frac{i \lambda  t}{N}\right)^{r}  \frac{1}{\prod_{i=0}^r n_i! } [H_{s_r},\cdots[H_{s_2},[H_{s_{1}},[H_{s_0},\mathcal{Q}]_{n_{0}}]_{n_{1}}\cdots ]]_{n_r}\nonumber\\
&\qquad =\sum_{s_1,\cdots,s_r =1}^L \frac{1}{r!}\left(\frac{i t r}{N}\right)^{r} [h_{s_r} H_{s_r},\cdots[h_{s_1} H_{s_1},\mathcal{Q}]]\cdots ], 
\end{align*}
After obtaining $p_{\vec{s}}$ and repeat this process every round until $N$ operators are drawn, the resultant operator is
\begin{align*}
\sum_{i,\vec{a}, \vec{b},\cdots \vec{z}}p^{\vec{y}}_{\vec{z}}\cdots p^{\vec{a}}_{\vec{b}} p_{\vec{a}}  \left(V_{\vec{z}} \cdots V_{\vec{b}} V_{\vec{a}} \right)^\dagger \mathcal{Q} \overbrace{ \left(V_{\vec{z}} \cdots V_{\vec{b}} V_{\vec{a}} V_{i}\right)}^{\text{ $N$ unitaries}}
\end{align*}
The Taylor expansion agrees with the exact evolution through order $r$. Under the at-most-linear error-accumulation assumption stated in Sec.~\ref{protocol}, accumulation over $M$ rounds gives the global error in Eq.~\eqref{qshift_global_error}.


At this point, let us mention that the probability distribution for $(N,r)$-qSHIFT algorithm is composed of $L^r$ variables, $p_{1\dots},\cdots p_{L\cdots L}$.   We need to validate that the number of independent equations constructed is also $L^r$. 
To prove that it is enough to consider $N=r$ case, since following our protocol, drawing $r(<N)$ operators case can be considered as another $r=N'$ case, hence we consider the $N=r$ case here. 

Following our protocol, we compare
\begin{align*}
&\sum_{\vec{s}} p_{r} V^\dagger_{\vec{s}}\mathcal{Q} V_{\vec{s}},\\
=&\sum_{\vec{s}} p_{\vec{s}} \mathcal{Q}\\
&+\sum_{\vec{s}} p_{\vec{s}} \left(\frac{i t \lambda}{N} \right) \sum_{s_i=1}^{L} [H_{s_i},\mathcal{Q}] \\
&+\sum_{\vec{s}} p_{\vec{s}} \left(\frac{i t \lambda}{N}\right)^2 \left(\frac{1}{2}\sum_{s_i=1}^{L}  [H_{s_i},\mathcal{Q}]_2 + \sum_{s_i,s_j=1}^L [H_{s_i},[H_{s_j},\mathcal{Q}]] \right)\\
&~\vdots\\
&+\sum_{\vec{s}} p_{\vec{s}} \left(\frac{i t \lambda}{N}\right)^N \left(\frac{1}{N!}\sum_{s_i=1}^{L}  [H_{s_i},\mathcal{Q}]_N +\frac{1}{(N-1)!} \sum_{s_i,s_j=1}^L \left([H_{s_i},[H_{s_j},\mathcal{Q}]_{N-1}] + [H_{s_i},[H_{s_j},\mathcal{Q}]]_{N-1}\right)  +\cdots\right.  \\
&\left.\quad\quad +\sum_{s_{i_1},\cdots s_{i_N}=1}^{L} [H_{i_1},[H_{i_2},\cdots, [H_{i_N},\mathcal{Q}]\cdots]] \right)\\
&+\mathcal{O}(t^{N+1}),\\
=&\sum_{\vec{s}} p_{\vec{s}} \mathcal{Q}\\
&+ \left(\frac{i t \lambda}{N} \right) \sum_{s_1 }   \sum_{s_2,\cdots s_N} \left(p_{s_1\cdots} + p_{\cdots s_1\cdots } +\cdots +p_{\cdots s_1}  \right) [H_{s_1},\mathcal{Q}]\\
&+ \left(\frac{i t \lambda}{N} \right)^2  \left( \sum_{s_1 }   \frac{1}{2} \sum_{s_2,\cdots s_N} \left(p_{s_1\cdots} + p_{\cdots s_1\cdots } +\cdots +p_{\cdots s_1}  \right) [H_{s_1},\mathcal{Q}]_2  + \sum_{s_1,s_2}   \sum_{s_3,\cdots s_N} \left(p_{s_1,s_2\cdots} + p_{\cdots s_1\cdots,s_2\cdots } +\cdots +p_{\cdots s_2,s_1}  \right) [H_{s_1},[H_{s_2},\mathcal{Q}] \right) \\
&~\vdots\\
&+ \left(\frac{i t \lambda}{N} \right)^N  \left( \sum_{s_1 }   \frac{1}{N!} \sum_{s_2,\cdots s_N} \left(p_{s_1\cdots} + p_{\cdots s_1\cdots } +\cdots +p_{\cdots s_1}  \right) [H_{s_1},\mathcal{Q}]_N +\cdots  +\sum_{s_1,s_2\cdots s_N} p_{s_1,s_2\cdots s_N} [H_{s_{i_1}},[H_{s_{i_2}},\cdots,[H_{s_{i_N}},\mathcal{Q}]\cdots]] \right) \\
&+\mathcal{O}(t^{N+1})
\end{align*}
with
\begin{align*}
&e^{i t H} \mathcal{Q} e^{-i t H} \\
=&\mathcal{Q}\\
&+\left(i t\right)  \sum_i h_i [H_i,\mathcal{Q}]\\
&+\frac{1}{2} (i t )^2 \sum_{ij} h_i h_j [H_i,[H_j,\mathcal{Q}]] \\
& \vdots\\
&+ \frac{1}{N!} (i t )^N \sum_{i_1\cdots i_N} h_{i_1} h_{i_2} \cdots  [H_{i_1},[H_{i_2}\cdots [H_{i_N},\mathcal{Q}]\cdots]]\\
&+\mathcal{O}(t^{N+1})
\end{align*}
order by order up to $\mathcal{O}(t^{r=N})$. 

Observe that similar equations repeatedly appear, involving terms like 
\begin{align*}
\sum_{s_2,\cdots s_N} \left(p_{s_1\cdots} + p_{\cdots s_1\cdots } +\cdots +p_{\cdots s_1}  \right),
\end{align*}
which arise at $\mathcal{O}(t^{s\geq1})$.
These redundancies allow us to perform systematic reduction analogous to Gaussian elimination at every order.
After carrying out the reduction process, the remaining independent equations are determined by the coefficient of nested commutators of the form $[H_{i_1},[H_{i_2},\cdots,[H_{i_N},\mathcal{Q}]]]$ without summation at $\mathcal{O}(t^{N})$. The number of such independent equations is therefore $L^N$, which completes the counting argument.

\subsection{Explicit construction of the adaptive signed coefficients}
\label{coefficient_derivation}

This section supplies the details of the construction summarized in
Sec.~\ref{coefficient_construction}. We first derive the history
coefficients. Suppose that, before round $p$,
\begin{align}
V_S^{(p-1)}
=V_{\eta_1}V_{\eta_2}\cdots V_{\eta_m},
\qquad
m=(p-1)r,
\end{align}
where $\eta_\ell$ is the index of the $\ell$-th sampled gate. For each
sampled gate $V_{\eta_\ell}$, introduce a nonnegative integer $n_\ell$
representing the Taylor order contributed by that gate. For example, for
a three-gate history and $q=2$, $(n_1,n_2,n_3)=(2,0,0)$ selects the
second-order term from $V_{\eta_1}$, whereas $(n_1,n_2,n_3)=(1,0,1)$
selects first-order terms from $V_{\eta_1}$ and $V_{\eta_3}$. In
general, a contribution of total order $q$ is specified by
$n_1,\ldots,n_m$ satisfying $n_1+\cdots+n_m=q$. Therefore, for
$\vec{ a}=(a_1,\ldots,a_q)$ with $0\leq q\leq r$,
\begin{align}
S_{a_1\cdots a_q}^{(p-1)}
=
\sum_{\substack{
n_1,\ldots,n_m\geq 0\\
n_1+\cdots+n_m=q\\
(a_1,\ldots,a_q)
=(\eta_m^{n_m},\ldots,\eta_1^{n_1})
}}
\frac{1}{n_1!\cdots n_m!}.
\label{history_combinatorial}
\end{align}
Here, $\eta_\ell^{n_\ell}$ denotes $n_\ell$ consecutive copies of
$\eta_\ell$. The reversed order arises from the nested-commutator order
under conjugation.

For efficient evaluation, the same coefficients can be generated
incrementally. Define the truncated contribution associated with a gate
$V_j$ by
\begin{align}
E_j
=
\mathbf 1+(j)+\frac{(j,j)}{2!}
+\cdots+
\frac{(\underbrace{j,\ldots,j}_{r})}{r!},
\label{single_gate_series}
\end{align}
where $\mathbf 1$ denotes the empty-sequence expansion. Using the
concatenation convolution in Eq.~\eqref{sequence_convolution}, the
complete history expansion is
\begin{align}
S^{(p-1)}
=
E_{\eta_m}*E_{\eta_{m-1}}*\cdots*E_{\eta_1},
\label{history_series}
\end{align}
with all sequences longer than $r$ discarded. Thus, when $V_j$ is
appended to the circuit, the history coefficients can be updated as
\begin{align}
S\leftarrow E_j*S
\label{history_update}
\end{align}
and truncated again at sequence length $r$.

The triangular matching equation in Sec.~\ref{coefficient_construction}
can also be written formally as
\begin{align}
P^{(p)}
=
T^{(p)}*
\left(S^{(p-1)}\right)^{-1},
\label{formal_history_inverse}
\end{align}
where the inverse is defined with respect to concatenation convolution.
Writing $S^{(p-1)}=\mathbf 1+\overline S^{(p-1)}$, the inverse through
sequence length $r$ is
\begin{align}
\left(S^{(p-1)}\right)^{-1}
=
\sum_{\nu=0}^{r}(-1)^\nu
\left(\overline S^{(p-1)}\right)^{*\nu},
\label{truncated_history_inverse}
\end{align}
where all sequences longer than $r$ are discarded.

We next construct the fixed sparse map $\mathcal B_r$ in
Eq.~\eqref{sol}. For a length-$r$ commutator-index sequence
$\vec{ a}$, let $\Pi_{\mathrm{const}}(\vec{ a})$ denote the
set of contiguous partitions whose blocks contain only repeated copies
of a single index. For example, the valid partitions of $(i,i,j)$ are
$[i][i][j]$ and $[ii][j]$, whereas $[i][ij]$ is excluded. For
$\pi\in\Pi_{\mathrm{const}}(\vec{ a})$, let
$\vec{ a}/\pi$ be the index sequence obtained by collapsing every
block to its common index, and define
\begin{align}
\kappa(\pi)
=
\prod_{b\in\pi}
\frac{(-1)^{|b|-1}}{|b|},
\label{partition_weight}
\end{align}
where $|b|$ is the number of entries in block $b$. The action of
$\mathcal B_r$ is
\begin{align}
\left[\mathcal B_rP^{(p)}\right]_{\vec{ a}}
=
\sum_{\pi\in\Pi_{\mathrm{const}}(\vec{ a})}
\kappa(\pi)P_{\vec{ a}/\pi}^{(p)}.
\label{sequence_signed_coefficient}
\end{align}
For fixed $L$ and $r$, this transformation can be constructed once and
reused in every round.

As a consistency check, for $r=2$,
\begin{align}
P_i^{(p)}
&=T_i^{(p)}-S_i^{(p-1)},\\
P_{ij}^{(p)}
&=T_{ij}^{(p)}-S_{ij}^{(p-1)}
-P_i^{(p)}S_j^{(p-1)}.
\label{r2_sequence_recurrence}
\end{align}
The constant-block transformation gives
\begin{align}
\left[\mathcal B_2P^{(p)}\right]_{ii}
&=P_{ii}^{(p)}-\frac{1}{2}P_i^{(p)},\\
\left[\mathcal B_2P^{(p)}\right]_{ij}
&=P_{ij}^{(p)},
\qquad i\neq j.
\label{r2_sequence_solution}
\end{align}
Using Eq.~\eqref{target_sequence_coefficient}, these become
\begin{align}
\left[\mathcal B_2P^{(p)}\right]_{ii}
={}&
\frac{1}{2}
\left(\frac{2ph_i}{\lambda}\right)^2
-S_{ii}^{(p-1)}
-
\left(\frac{2ph_i}{\lambda}-S_i^{(p-1)}\right)
S_i^{(p-1)}
-
\frac{1}{2}
\left(\frac{2ph_i}{\lambda}-S_i^{(p-1)}\right),
\label{r2_diagonal_solution}\\
\left[\mathcal B_2P^{(p)}\right]_{ij}
={}&
\frac{2p^2h_ih_j}{\lambda^2}
-S_{ij}^{(p-1)}-
\left(\frac{2ph_i}{\lambda}-S_i^{(p-1)}\right)
S_j^{(p-1)}.
\label{r2_offdiagonal_solution}
\end{align}
Equation~\eqref{r2_offdiagonal_solution} applies for $i\neq j$. The
final factor is $S_j^{(p-1)}$, as required by the prefix--suffix
convolution.
Thus, explicit form of the solution of $r=2$ limit at $p$-th round is
\begin{align}
p_{ij}^{(p)} &= 2p^2 \frac{h_i h_j}{\lambda^2}- c_{ij} -\left(2p \frac{h_i}{\lambda}-n_j \right) - \delta_{ij} p \frac{h_{j}}{\lambda}\\
n_i &=	\sum_\ell \delta_{\eta_{\ell},i}, \quad c_{ij} =	\sum_{\ell<k} \delta_{\eta_{k},i}\delta_{\eta_{\ell},j} . 
\end{align}
where $n_i$ is the number of occurrence of gate $V_i$ and $c_{ij}$ is the number of ordered pairs $(V_i, V_j)$ where $V_j$ precedes $V_i$.

\subsection{Variance of qSHIFT}
\label{variance}
In this section, let us compute the variance of qSHIFT protocol with $N=r$ case first,  and the $M=N/r$ round case mentioned at the end of this section. Since the ensemble average of qSHIFT is written as $\langle \mathcal{Q}(t)\rangle_{qSHIFT}=  \sum_{\vec{s}} q_{\vec{s}} \left( Z(p_{\vec{s}})  \text{sign}(p_{\vec{s}}) \langle V^\dagger_{\vec{s}} \mathcal{Q} V_{\vec{s}} \rangle \right)$, the variance  is defined as follows,
\begin{align*}
\text{var}{\langle \mathcal{Q}\rangle_{qSHIFT}} &=  \sum_{\vec{s}} q_{\vec{s}} \left( Z(p_{\vec{s}})  \text{sign}(p_{\vec{s}}) \langle V^\dagger_{\vec{s}} \mathcal{Q} V_{\vec{s}} \rangle \right)^2-\left(\sum_{\vec{s}} q_{\vec{s}} \left( Z(p_{\vec{s}})  \text{sign}(p_{\vec{s}}) \langle V^\dagger_{\vec{s}} \mathcal{Q} V_{\vec{s}} \rangle \right)\right)^2.
\end{align*}

The first term is
\begin{align*}
&\sum_{\vec{s}} q_{\vec{s}} \left( Z(p_{\vec{s}})  \text{sign}(p_{\vec{s}}) \langle V^\dagger_{\vec{s}} \mathcal{Q} V_{\vec{s}} \rangle \right)^2\\
=&\sum_{\vec{s}} q_{\vec{s}} (Z(p_{\vec{s}})^2 \langle V^\dagger_{\vec{s}} \mathcal{Q} V_{\vec{s}} \rangle\\
=& \sum_{\vec{s} } q_{\vec{s}}  Z(p_{\vec{s}})^2 \left(\langle\mathcal{Q}  \rangle + \left(\frac{i t \lambda}{N} \right)  \langle \sum_{i=1}^{N} [H_{s_i},\mathcal{Q}]   \rangle +\mathcal{O}(t^2) \right)^2\\
=& \sum_{\vec{s}} q_{\vec{s}} Z(p_{\vec{s}})^2  \langle \mathcal{Q}\rangle^2   + \left(\frac{i t \lambda}{N} \right) \sum_{\vec{s}} q_{\vec{s}} Z(p_{\vec{s}})^2 \langle \sum_{i=1}^{N} [H_{s_i},\mathcal{Q}]   \rangle  + \mathcal{O}(t^2)\\
=& Z(p_{\vec{s}})^2 \langle \mathcal{Q}\rangle ^2+ \left(\frac{i t \lambda}{N} \right) \sum_{\vec{s}} q_{\vec{s}} Z(p_{\vec{s}})^2 \langle \sum_{i=1}^{N} [H_{s_i},\mathcal{Q}]   \rangle  + \mathcal{O}(t^2)
\end{align*}
and the second term is
\begin{align*}
&\left(\sum_{\vec{s}} q_{\vec{s}} \left( Z(p_{\vec{s}})  \text{sign}(p_{\vec{s}}) \langle V^\dagger_{\vec{s}} \mathcal{Q} V_{\vec{s}} \rangle \right)\right)^2\\
=&\left(\sum_{\vec{s}}p_{\vec{s}} \langle\psi |V^\dagger_{\vec{s}} \mathcal{Q} V_{\vec{s}}| \psi \rangle \right)^2\\
=&\left(\sum_{\vec{s}} p_{\vec{s}} \langle \mathcal{Q}  \rangle  + \left(\frac{i t \lambda}{N}\right) \sum_{\vec{s}} p_{\vec{s}}\langle [H_{s_i},\mathcal{Q}]\rangle  + \mathcal{O}(t^2)\right)^2 \\
&=\langle\mathcal{Q} \rangle^2 + \left(\frac{i t \lambda}{N}\right) \langle \mathcal{Q}\rangle \sum_{\vec{s}}p_{\vec{s} }\langle [H_{s_i},\mathcal{Q}] \rangle + \mathcal{O}(t^2)
\end{align*}

Thus, the variance of $(N,r=N)$-qSHIFT per round is
\begin{align*}
&\text{var}(\langle\mathcal{Q} \rangle_{\text{qSHIFT}})  =\left( Z(p_{\vec{s}})^2 -1\right) \langle \mathcal{Q}\rangle^2 + \mathcal{O}(t)
\end{align*}
and the sampling cost per round $N_{\text{qSHIFT,sampling}}$ for precision $\varepsilon_{qSHIFT}= \mathcal{O}\left(\frac{\lambda^{1+r} t^{1+r} }{N^{1+r}} \right)<1$ is given by
\begin{align*}
N_{\text{qSHIFT,sample}} &=\frac{\text{var}(\langle\mathcal{Q} \rangle_{\text{qSHIFT}})}{\varepsilon^2}=
\begin{cases}
\mathcal{O}\left( \frac{Z(p_{\vec{s}})^2 - 1} {\varepsilon^2} \langle\mathcal{Q} \rangle^2 \right), &\text{for $Z(p_{\vec{s}})>1$,}\\
\mathcal{O}\left( \frac{\lambda^2 t^2 }{\varepsilon^2}  \right), &\text{for $Z(p_{\vec{s}})=1$.} \\
\end{cases}
\end{align*}

We now derive an upper bound on the multi-round overhead $\mathbb{E}_{Q}\big[\prod_{p=1}^{M}Z_{p}^{2}\big]$. Since the trajectory probabilities satisfy $\sum_{\mathcal{H}_M}\prod_{p=1}^{M}q_{\vec s^{(p)}}=1$, any bound on $Z_{p}$ that holds uniformly over realized histories immediately yields 
\begin{equation}
\mathbb{E}_{Q}\Big[\prod_{p=1}^{M}Z_{p}^{2}\Big]
\;\le\;\prod_{p=1}^{M}\Big(\sup_{\mathcal H_{p-1}}Z_{p}\Big)^{2}.
\label{eq:ub-product}
\end{equation}
It therefore suffices to bound $Z_{p}$ round by round, uniformly over
the history.

For a coefficient array $A$ indexed by ordered sequences, define the
level-wise $\ell^{1}$ norm
$\|A\|_{q}\equiv\sum_{|\vec a|=q}|A_{\vec a}|$. The concatenation
convolution of Eq.~\eqref{sequence_convolution} is submultiplicative level by level,
\begin{equation}
\|A*B\|_{q}\;\le\;\sum_{k=0}^{q}\|A\|_{k}\,\|B\|_{q-k},
\label{eq:ub-submult}
\end{equation}
by the triangle inequality.

The norms of the history and target coefficients can be evaluated
exactly. All entries of $E_{j}$ in Eq.~\eqref{single_gate_series}, and hence of $S^{(p-1)}$ in
Eq.~\eqref{history_series}, are nonnegative, so $\|S^{(p-1)}\|_{q}$ is a plain sum, where summing Eq.~\eqref{history_combinatorial} over all sequences of length $q$ removes the
sequence-matching constraint, and the multinomial theorem gives
\begin{equation}
\|S^{(p-1)}\|_{q}
=\sum_{\substack{n_{1},\dots,n_{m}\ge 0\\ n_{1}+\cdots+n_{m}=q}}
\ \prod_{\ell=1}^{m}\frac{1}{n_{\ell}!}
=\frac{m^{q}}{q!},
\qquad m=(p-1)r.
\label{eq:ub-Snorm}
\end{equation}
Similarly, from Eq.~\eqref{target_sequence_coefficient},
\begin{equation}
\|T^{(p)}\|_{q}
=\frac{1}{q!}\Big(\frac{pr}{\lambda}\Big)^{\!q}
\sum_{a_{1},\dots,a_{q}=1}^{L}h_{a_{1}}\cdots h_{a_{q}}
=\frac{(pr)^{q}}{q!}.
\label{eq:ub-Tnorm}
\end{equation}
Both norms are independent of the realized history, of the individual
weights, and of $L$.

Taking absolute values in the triangular recursion of Eq.~\eqref{sequence_recurrence} and
using Eqs.~\eqref{eq:ub-submult}--\eqref{eq:ub-Tnorm} together with
$m\le rp$,
\begin{equation}
\|P^{(p)}\|_{q}\;\le\;\frac{2(rp)^{q}}{q!}
+\sum_{k=1}^{q-1}\|P^{(p)}\|_{k}\,\frac{(rp)^{q-k}}{(q-k)!}.
\label{eq:ub-Prec}
\end{equation}
Let $\pi_{q}$ solve Eq.~\eqref{eq:ub-Prec} with equality, so that
$\|P^{(p)}\|_{q}\le\pi_{q}$ by induction. The generating function
$\Phi(\zeta)=\sum_{q\ge1}\pi_{q}\zeta^{q}$ obeys
$\Phi=2(e^{rp\zeta}-1)+\Phi\,(e^{rp\zeta}-1)$, i.e.
\begin{equation}
\Phi(\zeta)=\frac{2\,(e^{rp\zeta}-1)}{2-e^{rp\zeta}}.
\label{eq:ub-Phi}
\end{equation}
Since all $\pi_{q}\ge0$, evaluating Eq.~\eqref{eq:ub-Phi} at
$\zeta_{0}=\ln\tau/(rp)$ for any $\tau\in(1,2)$ gives the term-wise
bound
\begin{equation}
\|P^{(p)}\|_{q}\;\le\;\pi_{q}\;\le\;
C_{\tau}\Big(\frac{rp}{\ln\tau}\Big)^{\!q},
\qquad
C_{\tau}=\frac{2(\tau-1)}{2-\tau}.
\label{eq:ub-Pbound}
\end{equation}

It remains to control the map $B_{r}$. From Eq.~(B9) and the triangle
inequality,
\begin{equation}
Z_{p}=\sum_{|\vec a|=r}\big|[B_{r}P^{(p)}]_{\vec a}\big|
\;\le\;\sum_{|\vec a|=r}\ \sum_{\pi\in\Pi_{\rm const}(\vec a)}
|\kappa(\pi)|\,\big|P^{(p)}_{\vec a/\pi}\big|.
\label{eq:ub-Br1}
\end{equation}
A pair $(\vec a,\pi)$ is uniquely specified by the composition
$(b_{1},\dots,b_{k})$ of $r$ formed by the block sizes together with
the collapsed sequence $\vec a/\pi$ of length $k$, and
$|\kappa(\pi)|=\prod_{j}1/b_{j}$. Since
$\sum_{b\ge1}u^{b}/b=-\ln(1-u)$, the sums over compositions into $k$
parts are generated by $\big(-\ln(1-u)\big)^{k}$, and
Eq.~\eqref{eq:ub-Br1} becomes
\begin{equation}
Z_{p}\;\le\;\sum_{k=1}^{r}
\Big[[u^{r}]\big(-\ln(1-u)\big)^{k}\Big]\,\|P^{(p)}\|_{k}
\;\le\;[u^{r}]\,\Phi\big(-\ln(1-u)\big)
=[u^{r}]\,
\frac{2\big[(1-u)^{-rp}-1\big]}{2-(1-u)^{-rp}},
\label{eq:ub-Br2}
\end{equation}
where $[u^{r}]$ denotes the coefficient extraction and the resummation
in the second inequality uses the nonnegativity of all coefficients.
The final series has nonnegative coefficients and converges wherever
$(1-u)^{-rp}<2$; evaluating it at $u_{0}=1-\tau^{-1/(rp)}$, where it
equals $C_{\tau}$, gives $[u^{r}](\cdots)\le C_{\tau}u_{0}^{-r}$.
Using $1-e^{-a}\ge a\,e^{-a}$ with $a=\ln\tau/(rp)$, together with
$r/(rp)\le1$, we arrive at the history-uniform bound
\begin{equation}
Z_{p}\;\le\;\tau\,C_{\tau}\Big(\frac{rp}{\ln\tau}\Big)^{\!r},
\qquad \tau\in(1,2).
\label{eq:ub-Zp}
\end{equation}

Inserting Eq.~\eqref{eq:ub-Zp} into Eq.~\eqref{eq:ub-product} with the
choice $\tau=2-1/r$, for which $C_{\tau}=2(r-1)$ and
$\ln\tau=\ln2+\mathcal{O}(1/r)$, and using Stirling's formula,
\begin{align}
\ln\mathbb{E}_{Q}\Big[\prod_{p=1}^{M}Z_{p}^{2}\Big]
&\;\le\;2\sum_{p=1}^{M}
\Big[r\ln\frac{rp}{\ln\tau}+\ln(\tau C_{\tau})\Big]
\nonumber\\
&\;=\;2rM\ln M-2rM+2rM\ln\frac{r}{\ln2}
+\mathcal{O}\big(M\ln r+r\ln M\big),
\label{eq:ub-final-log}
\end{align}
that is,
\begin{equation}
\mathbb{E}_{Q}\Big[\prod_{p=1}^{M}Z_{p}^{2}\Big]
\;\le\;\big(M!\big)^{2r}
\Big(\frac{r}{\ln2}\Big)^{\!2rM}
e^{\mathcal{O}(M\ln r+r\ln M)}
\;=\;\big(M!\big)^{2r}\,e^{\mathcal{O}_{r}(M)}.
\label{eq:ub-final}
\end{equation}
The bound holds for every weight profile $\{h_{i}\}$ and every $L$;
moreover, since the matching conditions equate coefficients of powers
of $x=i\lambda t/N$, the signed coefficients $p^{(p)}_{\vec s}$, and
hence $Z_{p}$ and the overhead, are independent of $t$, $\lambda$ and
$N$. A matching lower bound with the same leading term $2rM\ln M$ is
attained already by the single trajectory that samples the
smallest-weight operator in every round, so the exponent beyond the
factorial in Eq.~\eqref{eq:ub-final} is $\Theta_{r}(M)$; this is the
sampling-complexity entry quoted in Table~\ref{gatenumber}.

%
%
%

\section{Examples with details of calculations}
\label{example}
In this section, let us provide explicit examples for various $N$ and $r$, which depend on the machine condition, when it comes to the implementation of our proposal.
\subsection{($L=2$,$N=2$, $r=1$)}
Consider 
\begin{align*}
H =h_1 H_1 + h_2 H_2.
\end{align*}
 Following our protocol, the drawing set is
\begin{align*}
\mathcal{V} = \{V_1= e^{-i \frac{t\lambda }{2}H_1},V_2=e^{-i \frac{t\lambda }{2}H_2}\}.
\end{align*}
At the first round, we draw $V_i$ with probability $p_i =\frac{h_i}{\lambda}$. Next, for the drawn operator $V_i$, we determine the probability for the next drawing. To set the probability, we compare two operators,  
\begin{align*}
&\sum_{j=1}^2 p^{(i)}_jV^\dagger_jV^\dagger_i \mathcal{Q} V_i V_j, \\
=& \sum_{j=1}^2 p^{(i)}_j \mathcal{Q}  + \sum_{j=1}^2 \left( \frac{i t \lambda}{2} \right) p^{(i)}_j \left( [H_i,\mathcal{Q}] + [H_j,\mathcal{Q}]\right) +\mathcal{O}(t^2),\\
=& \sum_{j=1}^2 p^{(i)}_j \mathcal{Q}  +\left( \frac{i t \lambda}{2} \right) [H_i,\mathcal{Q}]   + \sum_{j=1}^2 \left( \frac{i t \lambda}{2} \right) p^{(i)}_j [H_j,\mathcal{Q}] +\mathcal{O}(t^2),\\
=& \sum_{j=1}^2 p^{(i)}_j \mathcal{Q}  +  \left( i t \right)  \Big(\frac{\lambda}{2} \;  (p^{(i)}_{j=i}+1) [H_i ,\mathcal{Q}] + \frac{\lambda}{2}p^{(i)}_{j\neq i} [H_{j\neq i},\mathcal{Q}]\Big)+\mathcal{O}(t^2),
\end{align*}
and
\begin{align*}
&e^{i t H} \mathcal{Q} e^{-i t H},\\
=& \mathcal{Q} +\sum_{i=1}^2 \left( i t\right) [h_i H_i,\mathcal{Q}]+\mathcal{O}(t^2),
\end{align*}
order by order. 

From the comparison, the obtained equations are 
\begin{align*}
&\sum_{j=1}^{2} p^{(i)}_j=1, \quad \frac{\lambda}{2}( p^{(i)}_{j=i}+1) =h_i ,\quad \frac{\lambda}{2} p^{(i)}_{j\neq i} =h_j,\\
\Rightarrow& p^{(i)}_{j=i} = \frac{h_i-h_j}{\lambda}, ~p^{(i)}_{j\neq i} = \frac{2h_j}{\lambda}, 
\end{align*}
Thus, the total expectation value is
\begin{align*}
p^{(1)}_1 p_1 \left(V_1 V_1  \right)^\dagger \mathcal{Q} V_1 V_1 + p^{(1)}_2 p_1 \left(V_2 V_1  \right)^\dagger \mathcal{Q} V_1 V_2 +p^{(2)}_1 p_2 \left(V_1 V_2  \right)^\dagger \mathcal{Q} V_2 V_1 +p^{(2)}_2 p_2 \left(V_2 V_2  \right)^\dagger \mathcal{Q} V_2 V_2,
\end{align*}
and its associated error is $\mathcal{O}(t^2)$.

\subsection{($L=2$,$N=2$, $r=2$)}
Next, under the same setting, we consider more drawing at once. Compare 
\begin{align*}
&\sum_{ij} p_{ij} V^\dagger_j V^\dagger_i \mathcal{Q} V_i V_j,\\
&=\sum_{ij} p_{ij} \mathcal{Q}, \\
&+\sum_{ij} p_{ij} \left(\frac{i t \lambda}{2}\right) \left([H_{i},\mathcal{Q}] + [H_{j},\mathcal{Q}]\right),\\
&+\sum_{ij} p_{ij} \frac{1}{2}\left(\frac{i t \lambda}{2}\right)^2 \left([H_{i},\mathcal{Q}]_2 + [H_{j},\mathcal{Q}]_2 + 2 [H_{j},[H_{i},\mathcal{Q}]]\right),\\
&+\mathcal{O}(t^3)
\end{align*}
to
\begin{align*}
&e^{i t H} \mathcal{Q} e^{-i t H},\\
=& \mathcal{Q} +\sum_{i=1}^2 \left( i t\right) [h_i H_i,\mathcal{Q}]+\sum_{i=1}^2 \frac{1}{2} \left( i t\right)^2 [h_j H_j ,[h_i H_i,\mathcal{Q}]]+\mathcal{O}(t^3),
\end{align*}
which gives
\begin{align*}
&\sum_{ij} p_{ij}=1, \quad  \lambda p_{11} +\frac{\lambda}{2} \left(p_{21}+p_{12} \right) = h_1, \quad \lambda p_{22} +\frac{\lambda}{2} \left(p_{21}+p_{12} \right) = h_2, \quad 2\left(\frac{\lambda}{2}\right)^2 p_{12} =2 \left(\frac{\lambda}{2}\right)^2 p_{21} = h_1 h_2.
\end{align*}

The solution of the above equation is
\begin{align*}
p_{11} = \frac{h_1(h_1-h_2)}{\lambda^2},\quad p_{12} =p_{21}= \frac{2h_1 h_2 }{\lambda^2} , \quad p_{22} = \frac{h_2 (h_2-h_1)}{\lambda^2}.
\end{align*}

\subsection{($L=2$,$N=3$, $r=2$)}
As another example, we consider $(L=2,N=3, r=2)$ case. We draw the first operator $V_i$ with probability $p_i$. To determine the probability of drawing the next two operators, consider
\begin{align*}
&\sum_{a,b} p^{(i)}_{ab} V^\dagger_{ab} V^\dagger_i \mathcal{Q} V_i V_{ab},\\
=&\sum_{a,b} p^{(i)}_{ab} \mathcal{Q} \\
& + \sum_{a,b} p^{(i)}_{ab} \left( \frac{i t \lambda}{3} \right) \left([ H_{i},\mathcal{Q}]  + [H_{a},\mathcal{Q}] +[H_{b}, \mathcal{Q}] \right) \\
& + \sum_{a,b} p^{(i)}_{ab} \frac{1}{2}\left( \frac{i t \lambda}{3} \right)^2 \left([ H_{i},\mathcal{Q}]_2  + [H_{a},\mathcal{Q}]_2 +[H_{b}, \mathcal{Q}]_2  +2  [H_a, [H_i,\mathcal{Q}]] +  2  [H_b, [H_i,\mathcal{Q}]] +  2  [H_a, [H_b,\mathcal{Q}]]  \right) \\
&+ \mathcal{O}(t^3),\\
=&\sum_{a,b} p_{ab} \mathcal{Q}\\ 
&+ \left( \frac{i t \lambda}{3}\right) \left( [H_{i},\mathcal{Q}] +\left(\sum_j \left(p_{i j} + p_{j i}\right)\right)[H_{i},\mathcal{Q}] +\left(\sum_j \left(p_{\bar{i} j} + p_{j \bar{i}}\right)\right)[H_{\bar{i}},\mathcal{Q}] \right)\\
&+ \frac{1}{2}\left( \frac{i t \lambda}{3}\right)^2 \left([H_i ,\mathcal{Q}]_2  +\left(\sum_j \left(p_{i j} + p_{j i}\right)\right)[H_{i},\mathcal{Q}]_2 +\left( \sum_j \left(p_{\bar{i} j} + p_{j \bar{i}}\right)\right)[H_{\bar{i}},\mathcal{Q}]_2 \right)\\
&+ \frac{1}{2}\left( \frac{i t \lambda}{3}\right)^2 \left( \left( 2\sum_j \left(p_{i j} + p_{j i}\right)\right) [H_i ,\mathcal{Q}]_2  + \left(2 \sum_j \left(p_{\bar{i} j} + p_{j \bar{i}}\right)\right) [H_{\bar{i}},[H_i ,\mathcal{Q}]]  \right)\\
&+ \frac{1}{2}\left( \frac{i t \lambda}{3}\right)^2 \left( 2 p_{ii} [H_i ,\mathcal{Q}]_2 + 2 p_{\bar{i}\bar{i} }  [H_{\bar{i}},\mathcal{Q}] + 2 p_{i \bar{i}} [H_i ,[H_{\bar{i}},\mathcal{Q}]] + 2 p_{\bar{i} i } [H_{\bar{i}} ,[H_{i},\mathcal{Q}]]  \right)\\
&+\mathcal{O}(t^3,\\
=&\sum_{a,b} p_{ab} \mathcal{Q}\\ 
&+ \left( \frac{i t \lambda}{3}\right) \left( \left(\sum_j \left(p_{i j} + p_{j i}\right)+1 \right)[H_{i},\mathcal{Q}] +\left(\sum_j \left(p_{\bar{i} j} + p_{j \bar{i}}\right)\right)[H_{\bar{i}},\mathcal{Q}] \right)\\
&+ \frac{1}{2}\left( \frac{i t \lambda}{3}\right)^2 \left( \left(\sum_j \left(p_{i j} + p_{j i}\right)+1 +      \left(2\sum_j \left(p_{i j} + p_{j i}\right)\right)  +2p_{ii}               \right)[H_{i},\mathcal{Q}]_2 +\left( \sum_j \left(p_{\bar{i} j} + p_{j \bar{i}}\right) +  2p_{\bar{i}\bar{i}}\right)[H_{\bar{i}},\mathcal{Q}]_2 \right)\\
&+ \frac{1}{2}\left( \frac{i t \lambda}{3}\right)^2 \left(  \left(2 \sum_j \left(p_{\bar{i} j} + p_{j \bar{i}}\right)+2p_{\bar{i}i} \right) [H_{\bar{i}},[H_i ,\mathcal{Q}]] +2p_{i\bar{i}} [H_{i},[H_{\bar{i}},\mathcal{Q}]]  \right)\\
&+\mathcal{O}(t^3),
\end{align*}
where $V_{ab} = V_a V_b$, $\bar{i}\neq i$. Direct comparison of this to the ideal operator gives,
\begin{align*}
&\sum_{a,b} p_{ab}^{(i)}=1,\\
&\left(\frac{\lambda}{3} \right) \left(\sum_j \left(p_{ij} + p_{ji}\right) + 1\right) = h_i, \\
&\left(\frac{\lambda}{3} \right) \sum_j \left(p_{\bar{i}j} + p_{j\bar{i}}\right)  = h_{\bar{i}} ,\\
&\left(\frac{\lambda}{3} \right)^2\left( \sum_j \left(p_{ij} + p_{ji}\right) + 1  + \left( 2 \sum_j \left(p_{ij} + p_{ji}\right) + 2p_{ii}\right)      \right)  = h^2_{{i}} ,\\
&\left(\frac{\lambda}{3} \right)^2\left( \sum_j \left(p_{\bar{i}j} + p_{j\bar{i}}\right)+ 2p_{\bar{i}\bar{i}}     \right)  = h^2_{\bar{i}}, \\
&\left(\frac{\lambda}{3} \right)^2 2p_{i\bar{i}} = h_i h_{\bar{i}},\\
&\left(\frac{\lambda}{3} \right)^2 \left( 2 \sum_j \left(p_{\bar{i}j} + p_{j\bar{i}}\right) + 2p_{\bar{i} i} \right) = h_i h_{\bar{i}}.
\end{align*}

Thus, the solution is
\begin{align*}
&p_{ii} = \frac{1}{2} \left( \frac{3}{\lambda}\right)^2 \frac{9 h_i^2+ 2\lambda^2-9\lambda h_i}{9}, \\
&p_{\bar{i}\bar{i}} = \frac{1}{2} \left( \frac{3}{\lambda}\right)^2 \frac{3 h_{\bar{i}}^2-\lambda h_{\bar{i}}}{3}, \\
&p_{i\bar{i}} =\frac{1}{2} \left( \frac{3}{\lambda}\right)^2 h_i h_{\bar{i}}\\
&p_{\bar{i}i} = \frac{1}{2} \left( \frac{3}{\lambda}\right)^2 \left( h_i h_{\bar{i}} - \frac{2}{3}\lambda h_{\bar{i}}\right)
\end{align*}

\subsection{($L=2$,$N=3$, $r=3$)}
As the last example, we provide the probability distribution following $(N=3,r=3)$-qSHIFT protocol with details of calculations. 
\begin{align*}
&\sum_{ijk} p_{ijk} (V_i V_j V_k)^\dagger \mathcal{Q} V_i V_j V_k,\\
=&\sum_{ijk} p_{ijk} \mathcal{Q} \\
&+ \sum_{ijk} p_{ijk} \left(\frac{i t\lambda  }{3}\right) \left( [H_i ,\mathcal{Q}] +[H_j ,\mathcal{Q}]+[H_k ,\mathcal{Q}]\right)\\
&+ \sum_{ijk} p_{ijk} \left(\frac{i t \lambda}{3} \right)^2\left(\frac{1}{2!}\left([H_{i},\mathcal{Q}]_2+[H_{j},\mathcal{Q}]_2+[H_{k},\mathcal{Q}]_2 \right) + [H_i,[H_j,\mathcal{Q}]] + [H_j,[H_k,\mathcal{Q}]] + [H_i,[H_k,\mathcal{Q}]] \right)\\
&+\sum_{ijk} p_{ijk} \left( \frac{i t \lambda}{3}\right)^3 \left(\frac{1}{3!} \left( [H_i,\mathcal{Q}]_3 +[H_j,\mathcal{Q}]_3 +[H_k,\mathcal{Q}]_3 \right)  + \frac{1}{2!}  \left([H_i,[H_j,\mathcal{Q}]_2]+[H_i,[H_k,\mathcal{Q}]_2]+[H_j,[H_k,\mathcal{Q}]_2] \right) \right.\\
&\left. \quad + \frac{1}{2!}  \left([H_i,[H_j,\mathcal{Q}]]_2+[H_i,[H_k,\mathcal{Q}]]_2+[H_j,[H_k,\mathcal{Q}]]_2 \right) + [H_i,[H_j,[H_k,\mathcal{Q}]]] \right) + \mathcal{O}(t^4),\\
=&\sum_{ijk} p_{ijk} Q \\
&+ \sum_{i}  \left(\frac{it \lambda}{3} \right)  \sum_{jk} \left(p_{ijk } + p_{jik}+ p_{jki}\right)  [H_i,\mathcal{Q}]\\
&+ \sum_{i} \left(\frac{i t \lambda}{3} \right)^2 \left( \frac{1}{2!}  \sum_{jk} \left(p_{ijk } + p_{jik}+ p_{jki}\right)   + \sum_{j} \left( p_{iij}+p_{iji}+p_{jii} \right)\right)[H_i,\mathcal{Q}]_2\\
&+ \sum_{i j (i\neq j)} \left(\frac{i t \lambda}{3} \right)^2   \left( \sum_{k} \left( p_{ijk}+p_{ikj}+p_{kij} \right)\right) [H_i,[H_{j},\mathcal{Q}]]\\
&+ \sum_{i} \left(\frac{i t \lambda}{3} \right)^3 \left( \frac{1}{3!}  \sum_{jk} \left(p_{ijk } + p_{jik}+ p_{jki}\right)   + \frac{2}{2!} \sum_{j} \left( p_{iij}+p_{iji}+p_{jii} \right)+ p_{iii}\right)[H_i,\mathcal{Q}]_3\\
&+\sum_{ij (i\neq j)}  \left(\frac{i t \lambda}{3} \right)^3 \left( \frac{1}{2!}\sum_{k} \left(p_{i j k}+p_{k i j}+p_{i k j} \right) + p_{ijj}\right) [H_{i},[H_j,\mathcal{Q}]_2]\\
&+\sum_{ij (i\neq j)}  \left(\frac{i t \lambda}{3} \right)^3 \left( \frac{1}{2!}\sum_{k} \left(p_{i j k}+p_{k i j}+p_{i k j} \right) + p_{iij}\right) [H_{i},[H_j,\mathcal{Q}]]_2 \\
&+ \sum_{j\neq i}  \left(\frac{i t \lambda}{3} \right)^3 p_{iji} [H_{i},[H_{j},[H_{i},\mathcal{Q}]]] + \mathcal{O}(t^4).
\end{align*}
By comparing to the ideal time-evolution operator order by order, we obtain
\begin{align*}
&\sum_{ijk} p_{ijk} =1,\\
\nonumber\\
&\frac{\lambda}{3} \sum_{jk} \left(p_{ijk}+ p_{jik}+p_{jki} \right) = h_i,\\
&\Rightarrow \sum_{jk} \left(p_{ijk}+ p_{jik}+p_{jki} \right) = \frac{3 h_i }{\lambda},\\
\nonumber\\
&\left(\frac{\lambda}{3}\right)^2\left(\sum_{jk} \left(p_{ijk} + p_{jik}+p_{jki} \right) + 2\sum_{j} \left(p_{iij}+ p_{iji}+ p_{jii} \right)\right) = h_i^2,\\
&\Rightarrow      \sum_{j} \left(p_{iij}+ p_{iji}+ p_{jii} \right) = \frac{1}{2} \left(\left(\frac{3 h_i}{\lambda}\right)^2-\frac{3h_i}{\lambda}\right),\\
\nonumber\\
&2\left(\frac{\lambda}{3}\right)^2\left( \sum_{k} \left(p_{ijk}+ p_{ikj}+ p_{kij} \right)\right) = h_i h_j,\\
&\Rightarrow  \sum_{k} \left(p_{ijk}+ p_{ikj}+ p_{kij} \right) = \frac{1}{2}\left(\frac{3h_i}{\lambda}\right)\left(\frac{ 3 h_j}{\lambda} \right),\\
\nonumber\\
&3! \left(\frac{ \lambda}{3} \right)^3 \left( \frac{1}{3!}  \sum_{jk} \left(p_{ijk } + p_{jik}+ p_{jki}\right)   + \frac{2}{2!} \sum_{j} \left( p_{iij}+p_{iji}+p_{jii} \right)+ P_{iii}\right)=h_i ^3,\\
&\Rightarrow   \frac{1}{3!} \left(\frac{3h_i }{\lambda}\right)  +    \frac{1}{2} \left( \left(\frac{3h_i}{\lambda} \right)^2 - \frac{3h_i}{\lambda} \right)+ P_{iii}=\frac{1}{3!}\left(\frac{3h_i}{\lambda}\right)^3, \\
\nonumber\\
& 3! \left(\frac{ \lambda}{3} \right)^3 \left( \frac{1}{2}\sum_{k} \left(p_{i j k}+p_{k i j}+p_{i k j} \right) + p_{ijj}\right)=h_i h_j^2,\\
&\Rightarrow  \frac{1}{4}  \left(\frac{3h_i}{\lambda}\right)\left(\frac{3h_j}{\lambda}\right)  + p_{ijj}  = \frac{1}{3!}\left(\frac{3h_i }{\lambda}\right) \left(\frac{3h_j }{\lambda}\right)^2, \\
\nonumber\\
& 3! \left(\frac{ \lambda}{3} \right)^3 \left( \frac{1}{2}\sum_{k} \left(p_{i j k}+p_{k i j}+p_{i k j} \right) + p_{iij}\right)=h_i^2 h_j,\\
&\Rightarrow \frac{1}{4} \left(\frac{3h_i}{\lambda} \right) \left(\frac{3h_j}{\lambda} \right) + p_{iij} =\frac{1}{3!}\left(\frac{3h_i}{\lambda} \right)^2 \left(\frac{3h_j}{\lambda} \right)  ,\\
\nonumber\\
& 3! \left(\frac{ \lambda}{3} \right)^3 p_{iji}=h_i h_j h_i,\\
&\Rightarrow P_{iji} = \frac{1}{3!} \left(\frac{3h_i}{\lambda} \right)^2 \left(\frac{3h_j}{\lambda} \right).
\end{align*}

The solution of the equations is 
\begin{align*}
p_{iii} &= \frac{1}{6} \left(\frac{3h_i}{\lambda}\right)^3-\frac{1}{2} \left(\frac{3h_i}{\lambda}\right)^2  + \frac{1}{3} \left(\frac{3h_i}{\lambda}\right),\\
p_{ijj} &= \frac{1}{6} \left(\frac{3h_i}{\lambda}\right)\left(\frac{3h_j}{\lambda}\right)^2 -\frac{1}{4} \left(\frac{3h_i}{\lambda}\right)\left(\frac{3h_j}{\lambda}\right), \\
p_{iij}&= \frac{1}{6} \left(\frac{3h_i}{\lambda}\right)^2 \left(\frac{3h_j}{\lambda}\right) - \frac{1}{4} \left(\frac{3h_i}{\lambda}\right)\left(\frac{3h_j}{\lambda}\right),\\
P_{iji}&= \frac{1}{6} \left(\frac{3h_i}{\lambda}\right)^2\left(\frac{3h_j}{\lambda}\right).
\end{align*}
\end{document}